\documentclass{aa}  

\usepackage{graphicx}
\usepackage[varg]{txfonts}
\usepackage[colorlinks=true,allcolors=blue]{hyperref}
\usepackage{orcidlink}
\usepackage{xcolor} 
\usepackage{float}
\usepackage{subcaption}
\usepackage{amsmath}
\usepackage{tabularx}
\usepackage{natbib}

\begin{document}

   \title{An X-ray flaring event and a variable soft X-ray excess in the Seyfert LCRS B040659.9$-$385922 as detected with eROSITA}

%   \subtitle{I. Overviewing the $\kappa$-mechanism}

   \author{S. Krishnan\orcidlink{0000-0001-5375-9131}\inst{1,2}\thanks{E-mail: saikruba@camk.edu.pl},
          A.G. Markowitz\orcidlink{0000-0002-2173-0673}\inst{1,3}, 
          M. Krumpe\inst{4},
          D. Homan\inst{4},
          R. Brogan\inst{4}, S. Haemmerich\orcidlink{0000-0002-1113-0041}\inst{5}, M. Gromadzki\orcidlink{0000-0002-1650-1518}\inst{6}, T. Saha\orcidlink{0000-0001-5647-3366}\inst{1},  M. Schramm\inst{7}, D.E. Reichart\inst{8}, H. Winkler\inst{9}, S. Waddell\inst{10}, J. Wilms\orcidlink{0000-0003-2065-5410}\inst{5}, A. Rau\inst{10},
          Z. Liu\inst{10}, and 
          I. Grotova\inst{10}
%          et al.\
          }
  \institute{
   Nicolaus Copernicus Astronomical Center, Polish Academy of Sciences, ul. Bartycka 18, 00-716 Warszawa, Poland
   \and 
   Inter-University Centre for Astronomy and Astrophysics, Post Bag 4, Ganeshkhind, Pune University Campus, Pune 411007, India
   \and
   University of California, San Diego, Center for Astrophysics and Space Sciences, MC 0424, La Jolla, CA, 92093-0424, USA
   \and
    Leibniz-Institut für Astrophysik Potsdam, An der Sternwarte 16, 14482 Potsdam, Germany
    \and
    Remeis Observatory \& Erlangen Centre for Astroparticle Physics, Friedrich-Alexander-Universität Erlangen-Nürnberg, Sternwartstr.\ 7, 96049 Bamberg, Germany
    \and
    Astronomical Observatory, University of Warsaw, Al. Ujazdowskie 4, PL-00-478 Warsaw, Poland
    \and
    Graduate School of Science and Engineering, Saitama University, 255 Shimo-Okubo, Sakura-ku, Saitama City, Saitama 338-8570, Japan
    \and Department of Physics and Astronomy, University of North Carolina at Chapel Hill, Campus Box 3255, Chapel Hill, NC 27599-3255, USA
\and
Department of Physics, University of Johannesburg, Kingsway, Auckland Park, Johannesburg, 2006, South Africa
\and 
Max-Planck-Institut f\"{u}r extraterrestrische Physik, Giessenbachstra{\ss}e 1, 85748 Garching bei M\"{u}nchen, Germany
  }

\authorrunning{S.\ Krishnan et al.}
\titlerunning{Soft X-ray Flaring in LCRS B040659.9$-$385922}
%   \date{Received September 15, 1996; accepted March 16, 1997}

% \abstract{}{}{}{}{} 
% 5 {} token are mandatory
 
  \abstract
  % context heading (optional)
  % {} leave it empty if necessary  
   {Extreme continuum variability in extragalactic nuclear sources can indicate extreme changes in accretion flows onto supermassive black holes.}
  % aims heading (mandatory)
{We explore the multiwavelength nature of a continuum flare in the Seyfert LCRS B040659.9$-$385922. The all-sky X-ray surveys conducted by the Spectrum Roentgen Gamma (SRG)/extended ROentgen Survey with an Imaging Telescope Array (eROSITA) showed that its X-ray flux increased by a factor of roughly five over six months, and concurrent optical photometric monitoring with the Asteroid Terrestrial-impact Last Alert System (ATLAS) showed a simultaneous increase.}
  % methods heading (mandatory)
{We complemented the eROSITA and ATLAS data by triggering a multiwavelength follow-up monitoring program (X-ray Multi-Mirror Mission: \textit{XMM-Newton}, Neutron star Interior Composition Explorer: NICER; optical spectroscopy) to study the evolution of the accretion disk, broad-line region, and X-ray corona. During the campaign, X-ray and optical continuum flux subsided over roughly six months. Our campaign includes two \textit{XMM-Newton} observations, one taken near the peak of this flare and the other taken when the flare had subsided. }
  % results heading (mandatory)
   {The soft X-ray excess in both \textit{XMM-Newton} observations was power law-like (distinctly nonthermal). Using a simple power law, we observed that the photon index of the soft excess varies from a steep value of $\Gamma \sim 2.7$ at the flare peak to a relatively flatter value of $\Gamma \sim 2.2$ as the flare subsided. We successfully modeled the broadband optical/UV/X-ray spectral energy distribution at both the flare peak and post-flare times with the AGNSED model,  incorporating thermal disk emission into the optical/UV and warm thermal Comptonization in the soft X-rays. The accretion rate falls by roughly 2.5, and the radius of the hot Comptonizing region increases from the flaring state to the post-flare state. Additionally, from the optical spectral observations, we find that the
broad \ion{He}{ii}~$\lambda$4686 emission line fades significantly as the optical/UV/X-ray continuum fades, which could indicate a substantial flare of disk emission above 54 eV. We also observed a redshifted broad component in the H${\beta}$ emission line that is present during the high flux state of the source and disappears in subsequent observations.}
  % conclusions heading (optional), leave it empty if necessary 
   {A sudden strong increase in the local accretion rate in this source manifested itself via an increase in accretion disk emission and in thermal Comptonized emission in the soft X-rays, which subsequently faded. The redshifted broad Balmer component could be associated with a
transient kinematic component distinct from that comprising the rest of the broad-line region.}


   \maketitle
%
%-------------------------------------------------------------------
\nolinenumbers 
\section{Introduction}

\renewcommand{\arraystretch}{1.30}
\begin{table*}
\caption{Observation log of all X-ray data.}             
\label{table:1}      
\centering          
\begin{tabular}{c  c c c  }     % 7 columns 
\hline\hline       
                      % To combine 4 columns into a single one 
Telescope/Instrument & Obs.\ Start  & Exposure Time$^a$ & $f_{0.2-5.0~{\rm keV}}^b$ \\
\hline                    
   eRASS1 & 11 February 2020 & 0.58 (0.31) &  $0.29^{+0.10}_{-0.06}$ \\  
   eRASS2 &     9 August 2020 & 0.52 (0.27) &   $1.21^{+0.16}_{-0.13}$  \\
   eRASS3 &     29 January 2021  & 0.38 (0.21) &   $1.16^{+0.15}_{-0.18}$  \\
   eRASS4 &   6 August 2021 & 0.39 (0.22) &  $5.19^{+0.32}_{-0.28}$   \\
  \textit{XMM-Newton} EPIC &   21 August 2021 &  26.40 (16.82, 24.15) &  $5.36^{+0.02}_{-0.05}$    \\  %--20:24:47
  NICER     & 24 August 2021 & 14.87 (7.42) & $5.83^{+0.08}_{-0.06}$ \\ %--23:45:54
   NICER     & 25 November 2021 & 21.84 (21.58) & $3.99^{+0.02}_{-0.25}$ \\  %--16:35:40
  eRASS5 &      31 January 2022  &  0.42 (0.20)&   $1.92^{+0.24}_{-0.21}$  \\
  \textit{XMM-Newton} EPIC &    10 March 2022 &   25.65 (17.06, 19.72) &  $2.14^{+0.02}_{-0.07}$  \\  %--01:49:25
 
\hline               
\end{tabular}
\tablefoot{$^{(a)}$Net exposure in kiloseconds. Values outside parentheses refer to exposure times
before any screening criteria are applied; values inside parentheses are good exposure times after screening.
For \textit{XMM-Newton}, the two good exposure times listed
refer to the pn and to each MOS, respectively.
In addition, for eRASS, the values outside
parentheses are times in the field of view only, that is, not corrected
for vignetting.  The values inside parentheses are corrected for
vignetting.
$^{(b)}$ Flux over 0.2--5.0 keV in units of $10^{-12}$~erg~ cm$^{-2}$~s$^{-1}$. All flux values are determined based on spectral fits in Sect.~\ref{sec:sec3}. 
%The eRASS exposure times are not corrected for  vignetting, and refer to time in the field of view only. 
}
 \label{tab:obs_tab1}
\end{table*}

An active galactic nucleus (AGN) is powered by an accreting supermassive black hole (SMBH) at the center of a galaxy. A common characteristic exhibited by persistently accreting AGNs is stochastic variability, which is present across all wavebands. Such variability is aperiodic, with fluctuations spanning over several orders of magnitude in temporal frequency (e.g., \citealt{2004ApJ...617..939M}). However, more extreme flux and spectral variability in X-ray and/or optical fluxes has been observed in both Seyferts and quasars over timescales of months to years (e.g., \citealt{2000A&A...355..485G}; \citealt{2020MNRAS.491.4925G}).
These drastic changes in the continuum flux are sometimes accompanied by significant changes in the optical spectra, specifically the fluxes of broad components of Balmer emission lines. 
Most cases are generally thought to be associated with strong variations in the global accretion rate onto the supermassive black hole (e.g., \citealt{2008AJ....135.2048T}; \citealt{2014ApJ...788...48S}; \citealt{2014ApJ...796..134D}; \citealt{2015ApJ...800..144L}; \citealt{2016MNRAS.455.1691R}). 
The community is still assessing the full range of physical mechanisms underlying the diversity of observed transient behaviors. Mechanisms proposed to explain the transient phenomena generally focus on structural changes in the accretion flow associated with significant changes in the accretion rate. Some models posit cooling or heating fronts, as in \citet{2018ApJ...864...27S}, \citet{2018MNRAS.480.3898N}, and \citet{2018MNRAS.480.4468R}. In other cases, impacts from streams associated with tidal disruption events (TDEs) may be relevant (e.g., \citealt{2020ApJ...898L...1R}). A few open questions associated with these events are how the various structural components of the accretion inflow and outflow respond -- in terms of geometry and emission characteristics -- to such major changes in accretion rate and if certain components are even present or absent in some cases.

As a review of the standard emission components in Seyfert AGN, the optical/UV emission is thought to originate from a geometrically thin, optically thick accretion disk that is approximated as a multicolor blackbody spectrum (\citealt{1973A&A....24..337S}; \citealt{1973blho.conf..343N}).
Due to absorption along the line of sight by the Galaxy as well as by the AGN host galaxy, 
%absorption and reprocessing by the BLR (\citealt{2012MNRAS.423..451L}), 
the peak of the thermal UV/EUV emission from this component is hard to detect. 
The X-ray spectrum of Seyfert galaxies typically has a cutoff power-law continuum with reflection features, low energy absorption, and often a soft X-ray excess.
The primary X-ray power-law continuum is believed to be produced due to the Comptonization of the optical-UV soft seed photons emitted thermally from the accretion disk (``cold phase'') by energetic electrons in a hot compact region called the corona. 
The corona typically has an electron temperature $k_{\rm B}T_{\rm e} \sim 100$ keV (``hot phase'') and is optically thin %where optical depth $\tau \sim 0.5$ 
(\citealt{1991ApJ...380L..51H}; \citealt{1994ApJ...432L..95H}; \citealt{1995ApJ...438L..63Z},\citeyear{1996MNRAS.283..193Z}). A part of this primary emission can be reprocessed by interacting with material close to the central black hole in the accretion disk (\citealt{1991MNRAS.249..352G}; \citealt{1991A&A...247...25M}) or with the distant dusty torus \citep[e.g.,][]{Murphy09},
%\citealt{1993ARA&A..31..473A}; \citealt{2004Natur.429...47J}; \citealt{2009MNRAS.394.1325R}) 
generating a reflection spectrum. The reflection components are composed of a Compton hump peaking at $\sim$ 20--30 keV and an Fe~K$\alpha$ fluorescent emission line around 6--7~keV, depending on the ionization state of iron. 
Additionally, the parsec-scale dusty torus absorbs the emission from the accretion disk and reemits it in the near- and mid-infrared at wavelengths longer than $\sim 1 {\mu}$m. Depending on whether the torus intersects the line of sight and depending on the morphology and viewing effects, the torus can obscure X-rays and give rise to silicon emission or absorption features in the mid-IR \citep[e.g.,][and references therein]{Hao07}. 
 
In the absence of torus absorption of the X-ray continuum, soft X-ray excess emission is typically observed below 1--2 keV (\citealt{1985MNRAS.217..105A}; \citealt{1985ApJ...297..633S}). 
The soft excess has been observed in more than 50\% of Seyfert~1 galaxies (\citealt{1984ApJ...281...90H}; \citealt{1989MNRAS.240..833T}). 
The origin of the soft excess is still not completely understood. 
A blackbody is generally excluded, as its characteristic temperature has been found to be independent of the black hole mass (\citealt{2004MNRAS.349L...7G}; \citealt{2009A&A...495..421B}). Preferred explanations include (i) a "warm Comptonization" region, wherein the seed disk photons are up-scattered in an optically thick ($\tau \sim$ 10--40) and warm ($k_{\rm B}T_{\rm e} \lesssim 1$ keV) plasma (e.g., \citealt{2004A&A...422...85P}; \citealt{2011A&A...534A..39M}; \citealt{2018A&A...611A..59P}), and (ii) blurred ionized reflection, wherein the soft excess is produced due to relativistic blurred reflection of the primary X-ray continuum in an ionized disk (e.g., \citealt{2006MNRAS.365.1067C}).

The X-ray and optical emission from AGNs is highly variable, not only in flux but also in broadband spectral shape. Multiwavelength monitoring, including searches for inter-band lags or leads, can yield insight into the nature, structure, and physical properties of various accretion flow components. For example, observations of optical/UV photons lagging behind the X-rays are consistent with irradiation of the disk by the X-ray emission from the compact corona and subsequent thermal reprocessing, where the lags are consistent with light travel times from corona to disk (e.g., \citealt{2003ApJ...584L..53U}; \citealt{2014MNRAS.444.1469M}; \citealt{2015ApJ...806..129E}; \citealt{2016ApJ...821...56F}). An alternate interpretation of these lags focuses on the broad-line region (BLR) emitting reprocessed continuum radiation \citep{2022MNRAS.509.2637N}. On the other hand, observations of X-rays lagging the optical/UV emission are sometimes attributed to inwardly propagating local mass accretion rate fluctuations (e.g., \citealt{2003MNRAS.343.1341S}; \citealt{2008ApJ...677..880M}).

   \begin{figure*}
   \centering
    \includegraphics[width=12cm]{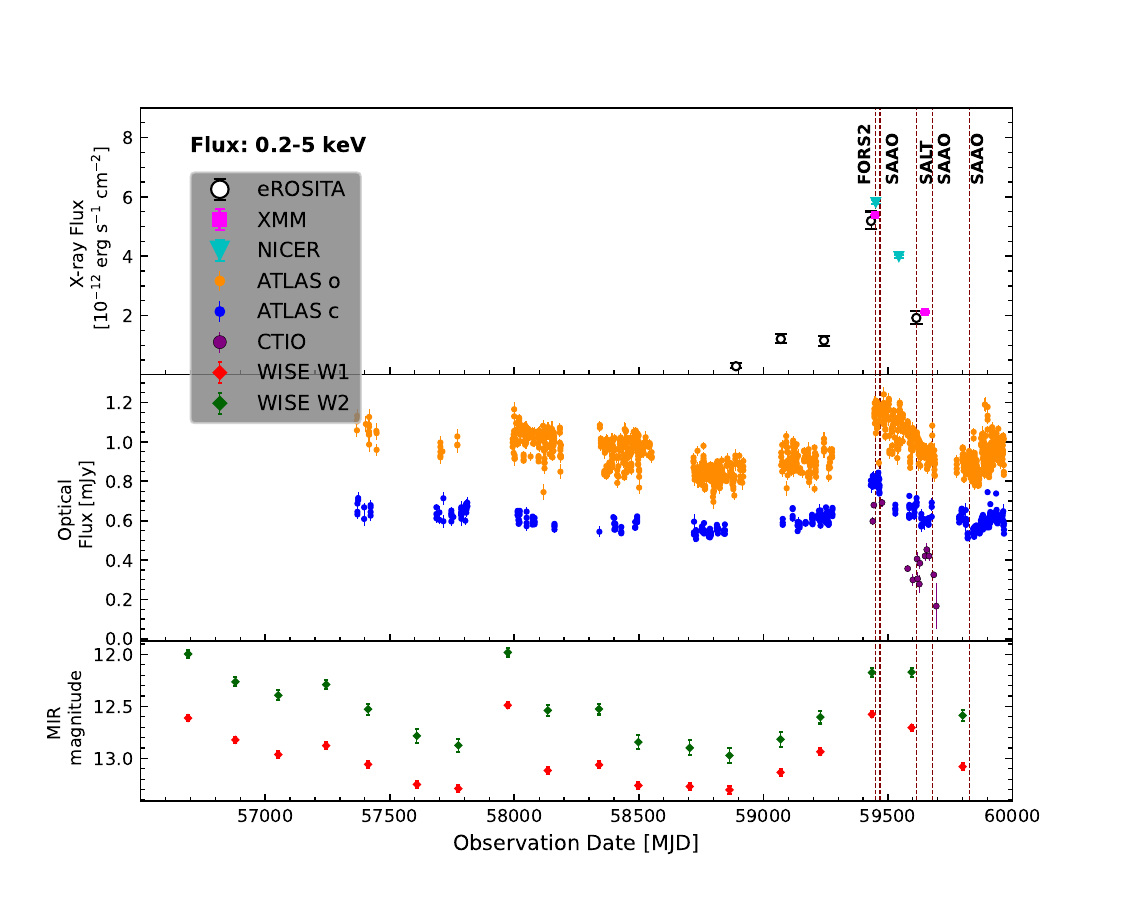}
      \caption{Multiwavelength light curves of J0408$-$38.  In the top panel, we plot the 0.2--5 keV X-ray flux from the \textit{XMM-Newton} observations, the five eRASS scans, and the two NICER observations. In the middle panel, we plot the optical light curves from the ATLAS photometric monitoring, showing data taken in the cyan (c, covering 420--650 nm) and orange (o, 560--820 nm) filters, as well as data from our B-band photometric monitoring at CTIO, shown in purple. In the bottom panel, we plot the MIR light curve from WISE's $W1$ and $W2$ bands. The source was flaring around the same time in multiple wavelengths. The epochs of optical spectroscopy are denoted by red vertical dotted lines.
              }
         \label{fig:obs_fig1}
   \end{figure*} 

The extended ROentgen Survey with an Imaging Telescope Array \citep[eROSITA;][]{2021A&A...647A...1P} is the soft X-ray instrument on the Russian Spectrum-Roentgen-Gamma \citep[SRG;][]{Sunyaev21} mission. It was developed by a German consortium led by the Max Planck Institute for extraterrestrial physics (MPE). The SRG was launched successfully on 13 July 2019 from the Baikonur cosmodrome, Kazakhstan, and it now orbits the Earth-Sun L2 position. The primary goals of eROSITA are to map the entire sky in X-rays and to detect 50,000--100,000 clusters of galaxies up to a redshift of $z \sim 1.3$ in order to track the cosmological evolution of large-scale structures. Additionally, eROSITA yields a sample of roughly $ 10^6$ AGN. Notably, eROSITA has already found several new highly variable extragalactic nuclear sources, including TDE candidates, changing-look AGNs, and other accretion-driven flaring sources (e.g., \citealt{2021MNRAS.508.3820S}; \citealt{2021A&A...647A...9M}; \citealt{2021MNRAS.502L..50S}; \citealt{2023A&A...672A.167H}; \citealt{2023A&A...669A..75L}).

In this paper, we discuss a flare in the Seyfert galaxy LCRS B040659.9$-$385922 (eRASSt\_J040846.1--385132; henceforth J0408$-$38) detected by eROSITA that shows significant X-ray flux variations between eROSITA's successive all-sky scans. 
We performed a multiwavelength follow-up campaign, making use of space-based telescopes such as the X-ray Multi-Mirror Mission (\textit{XMM-Newton}) and Neutron star Interior Composition Explorer (NICER) to cover the UV to X-ray wavebands as well as public optical survey data.
We determined that the source was flaring in the optical-UV and X-ray continua simultaneously.
As discussed below, we find that the spectral shape of the soft excess varied significantly between the X-ray observations taken during the flaring epoch and after the flaring of the source.
X-ray spectral fits and broadband spectral energy distribution (SED) fits indicate that the soft excess can be explained using a thermal Comptonization model. 
In addition, our optical spectral observations of the source reveal that the \ion{He}{ii} $\lambda$4686 emission line fades in intensity, roughly tracking the optical/UV/X-ray continuum. 

The paper is structured as follows. In Sect.~\ref{sec:sec2} we present our multiwavelength follow-up data sets and data reduction techniques. %, including public optical data. 
In Sect.~\ref{sec:sec3} we present our X-ray spectral analysis. In Sect.~\ref{sec:sec4} we present the results from modeling the broadband optical-UV to X-ray SED. In Sect.~\ref{sec:opt} we present the analysis from our optical spectroscopic follow-up observations. We discuss our results in Sect.~\ref{sec:sec6} and summarize our conclusions in Sect.~\ref{sec:sec7}.

All magnitudes quoted in this paper are Vega magnitudes.
All uncertainties for model parameter values determined from X-ray and broadband SED fits are 90\% confidence levels.

\section{Source detection, follow-up observations, and data reduction}
\label{sec:sec2}

\subsection{eROSITA source detection and data reduction}
The position of J0408$-$38 was scanned by eROSITA during each of the five eROSITA All-Sky Surveys \citep[eRASS;][]{2024A&A...682A..34M}. Data products were extracted using the eROSITA data analysis software eSASS version \texttt{eSASSuser 211214} (\citealt{2022A&A...661A...1B}). The eRASS tasks \texttt{evtool} and \texttt{srctool} were employed to handle the event data and extract the source products, respectively. Data processing version c020 was used. We combined data from all seven Telescope Modules. Source spectra were extracted using circular regions whose radii were varied in size, depending on the brightness of the source:
61$\arcsec$, 94$\arcsec$, 92$\arcsec$, 134$\arcsec$, and 78$\arcsec$, respectively, for eRASS1--5. Backgrounds were extracted from annular regions, with nearby point sources excised; background inner and outer annuli were  
136$\arcsec$--759$\arcsec$,
196$\arcsec$--1168$\arcsec$,
190$\arcsec$--1143$\arcsec$,
266$\arcsec$--1663$\arcsec$,
and 162$\arcsec$--969$\arcsec$, respectively.

The source was detected by eROSITA during its ignition phase: there was an increase in the X-ray flux as determined by comparison of successive eRASS scans. The flux increased by a factor of 4.5 over a period of six months from $f_{0.2-5.0}= 1.16^{+0.15}_{-0.18} \times 10^{-12}$~erg~cm$^{-2}$~s$^{-1}$ in eRASS3 (January 2021) to $f_{0.2-5.0}= 5.19^{+0.32}_{-0.28} \times 10^{-12}$~erg~cm$^{-2}$~s$^{-1}$ in eRASS4 (August 2021). Considering an earlier flux measurement of $f_{0.2-5.0} = 0.29^{+0.10}_{-0.06}\times 10^{-12}$~erg~cm$^{-2}$~s$^{-1}$  during eRASS 1 (February 2020), the flux increase was a factor of 18 over 18 months. Then, the flux dropped back toward the lower flux level, decreasing a factor of almost 3 by eRASS5 (January 2022). The fluxes from eRASS along with the other X-ray flux values from our follow-up program are listed in Table~\ref{tab:obs_tab1} and plotted in Fig.~\ref{fig:obs_fig1}. The X-ray position of the object is at coordinates (RA, Dec) = ($62.19204^{\circ}$, $-38.85875^{\circ}$) with a positional uncertainty of $\sim$$1.0\arcsec$ (includes both statistical and systematic uncertainties); the source is thus named eRASSt\_J040846.1$-$385132. The closest counterpart in the optical and IR bands is the Seyfert galaxy WISEA J040846.10$-$385131.0 $=$ LCRS B040659.9$-$385922, with optical coordinates of (RA, Dec) = ($62.19208^{\circ}$, $-38.85861^{\circ}$), at a redshift of $z=0.0574$ \citep{Shectman96}. The optical to X-ray offset is about $0\farcs36$. We obtained X-ray follow-up observations using \textit{XMM-Newton} and NICER during the flaring state and during the declining flux state of the source, as described below.

%d

\subsection{XMM-Newton data reduction}

The first \textit{XMM-Newton} observation (ObsID 0862770501) was taken during J0408$-$28's flaring state, on 21 August 2021 (XMM1), with a 26.40 ks exposure. By March 2022, the X-ray flux of the source had decreased significantly, back to roughly the same flux levels before the flaring one year earlier. 
\textit{XMM-Newton} thus observed the source again on 10 March 2022 (ObsID 0903990901), with a 25.65 ks exposure (XMM2).
In XMM1, the EPIC pn and MOS camera were operated in Small Window and Partial Window imaging modes respectively. Both cameras were operated in the Large Window imaging mode during XMM2. The data were processed using \textit{XMM-Newton} science analysis system (SAS) version 19.1.0, applying the latest calibration files and using HEASOFT (v.6.28). We followed the standard procedure for processing the EPIC pn and MOS data using the \texttt{epproc} and \texttt{emproc} tasks within SAS, respectively. Then we removed the time bins having high background by creating a Good Time Interval (GTI) file using the task \texttt{tabgtigen}. Final good exposure times after screening for XMM1 were 16.8 and 24.2 ks for pn and for each MOS, respectively. Final exposure times for XMM2 were 17.1 and 19.7~ks for pn and each MOS, respectively. The source and background spectra were extracted using a circular region with radius  $35\arcsec$ for both observations. The data do not show any significant pile-up as indicated by the SAS task \texttt{epatplot}. The spectra were rebinned with \texttt{grppha} with a minimum of 20 counts per bin to ensure the use of $\chi^2$ statistics.  For the pn, we extracted single (pattern 0) and double (patterns 1–4) spectra separately (hereafter pn0 and pn14). Redistribution matrices and ancillary response files were generated with the SAS tasks \texttt{rmfgen} and \texttt{arfgen}, respectively. 

In both observations, \textit{XMM-Newton}'s Optical Monitor imaged J0408$-$38 using the UV M2 filter (effective wavelength 2310~\AA). Data were taken in both "image" and "fast" modes and reduced using the XMM\_SAS routines \texttt{omichain} and \texttt{omfchain}, respectively. These routines apply all necessary calibrations (including flat-fielding and correcting exposures for dead time), and they perform point-source aperture photometry. We obtained AB magnitudes of $17.464 \pm 0.008$ for XMM1 and $18.906 \pm 0.022$ for XMM2, but these values are not yet corrected for Galactic extinction nor for systematic uncertainties. We corrected for Galactic extinction using E(B$-$V) = 0.0081 at the sky position of J0408$-$38 (\citealt{1998ApJ...500..525S}), and use $R \equiv A_{\rm V} / E(B-V) = 3.1$, which yields $A_{\rm V} = 0.025$~mag.  We use the Galactic extinction curve of \citet{1989ApJ...345..245C} to estimate the extinction near M2 as $A_{\rm M2} = 2.1 \times A_{\rm V} = 0.053$ mag. The corrected AB magnitudes are thus $17.411 \pm 0.008$ for XMM1 and $18.853 \pm 0.022$  for XMM2.  The corresponding extinction-corrected flux densities are  $(22.125\pm0.159) \times10^{-16}$~erg~cm$^{-2}$~s$^{-1}$ \AA$^{-1}$ ($0.394\pm0.003$~mJy) for XMM1, and  $(5.861\pm0.114) \times10^{-16}$~erg~cm$^{-2}$~s$^{-1}$ \AA$^{-1}$ ($0.104\pm0.002$~mJy) for XMM2.

\subsection{NICER data reduction}

NICER observed J0408$-$38 in two campaigns, as summarized in Table~\ref{tab:obs_tab1}. The first campaign consisted of ten observations (\-ObsIDs 4595040101--4595040110), starting
24 August 2021 (three days after the first \textit{XMM-Newton} observation) and lasting until 21 September 2021, and totaling 14.87 ks of exposure.
% with a weighted midpoint of 6 September 2021. (MJD 59463)
% " \- " tells latex to prevent line breaking; without it, latex was doing Ob-  sIDs
The second campaign covered four observations (ObsIDs 4595040201-4595040204), spanning 25 to 30 November 2021, and totaling 21.84 ks in exposure. 
% weighted midpoint of 27 Nov 2021. (MJD 59545)
Extraction of source spectra, background spectra, response files, and ancillary response files all followed standard procedures using the NICER data reduction pipeline. Background spectra were based on the 3C50 background estimator. Counts from the two noisy detectors, 14 and 34, were not excluded, although we verified that their inclusion or exclusion did not impact results.  To screen out periods of extreme optical loading impacting energies above 0.3~keV, we rejected time intervals when the detector undershoot rate -- detector resets caused when incoming optical photons trigger a release of the accumulated charge \citep{Remillard2002} -- exceeded 150 ct s$^{-1}$. We filtered against periods of high-energy particle rates by screening out time intervals when the overshoot rate exceeded 1.5 ct s$^{-1}$.  We also applied a background hard band count rate cut of 0.5 ct s$^{-1}$. Good exposure times after screening were 7.42 and 21.58 ks
for the August--September and November observations, respectively.
In all spectral fits described in Appendix \ref{appendix:nicer}, we ignore the data below 0.4~keV due to the effects of optical loading affecting the softest energy bins, and we can obtain good model constraints fitting up to 5~keV. Spectra are grouped to a minimum of 20 counts per bin.

\subsection{Optical and IR photometric data}

As part of this study, we use the publicly available photometric data from the Asteroid Terrestrial-impact Last Alert System (ATLAS) survey, which were obtained by running forced photometry on the reduced images (\citealt{2018ApJ...867..105T}; \citealt{2020PASP..132h5002S}). The observations from ATLAS are done using the orange (o, 560--820 nm) and cyan (c, 420--650 nm) filters. The resulting photometrically calibrated fluxes are reported in Fig.~\ref{fig:obs_fig1}. 

We obtained infrared photometry taken using the WISE telescope’s NEOWISE-R project, using the NASA IRSA archives. The WISE bands W1 and W2 are centered on 3.368 and 4.618 microns, respectively. We use the fluxes generated by the automated forced photometry pipeline. We binned the data to a six-month timescale, averaging both fluxes and uncertainties. As shown in Fig.~\ref{fig:obs_fig1}, the optical and IR light curves show qualitatively similar variability trends to the X-ray data. The object exhibits a sudden increase and then decrease in flux at all continuum wavelengths probed, with both the increase and decrease lasting approximately six months. 

We also obtained 14 B-band observations of J0408$-$38 at the 0.4-m
  PROMPT6 telescope at Cerro Tololo Inter-American Observatory (CTIO),
  operated as part of Skynet, during MJD 59439--59694.  All images
  were reduced following standard procedures, which included bias
  correction and flat-fielding.  Aperture photometry followed standard
  procedures using circular extraction regions for the source and
  source-centered annular regions for the background.  The resulting
  Vega B-band magnitudes were converted into mJy are
  plotted in Fig.~\ref{fig:obs_fig1}; they are not
  corrected for Galactic reddening.

\subsection{Optical spectroscopic observations}

We obtained a total of five optical spectra of the counterpart of J0408$-$38 as part of our multiwavelength follow-up campaign. A single spectrum was taken on 23 August 2021 (MJD 59449), during the flaring phase, using the FORS2 instrument \citep{APPENZELLER_1998}, installed on the UT1 telescope of the ESO 8.2~m Very Large Telescope in Chile. The target was observed with a 1.3-arc-second slit, using three grism+filter configurations: G14OOV (spectral resolution R=2100; 1000~s exposure), G300V + GG435+81 (R=440, 600~s exposure), and G300I + OG590+32 (R=660, 400~s exposure). We combined these exposures to produce a single high-quality spectrum.

Three spectra were taken with the SpUpNIC spectrograph \citep{CRAUSE_2019} at the 1.9~m telescope, also located at the SAAO. One observation occurred on 11 September 2021 (MJD 59468), during the flaring phase. Two additional observations were made as the flare faded, on 11 April 2022 (MJD 59680) and 6 September 2022 (MJD 59828). We made use of grating 7 at an angle of 16.5 degrees, resulting in wavelength coverage of 3500--8500~$\AA$ and an wavelength resolution of $R=500$. Each observation had a total integration time of approximately 2400~s. All spectra were reduced using standard bias corrections, flat-fielding, and wavelength calibration using arc-lamp spectra taken during the same night. Spectrophotometric calibrations were performed using a standard star taken on the night in the case of the VLT and SAAO 1.9~m observations, and a standard star taken within the past year in the case of the SALT spectrum.

Finally, a further spectrum was taken during the declining phase of the flare on 6 February 2022 (MJD 59616) using the Robert Stobie Spectrograph \citep[RSS;][]{BURGH_2003,KOBULNICKY_2003} mounted on the 10~m Southern African Large Telescope \citep[SALT;][]{BUCKLEY_2006}, situated at the South African Astronomical Observatory (SAAO) in the Northern Cape province of South Africa. The observation used the pg0900 grating. 
We used two camera setups. One spectrum was taken using a grating angle of 15.875$^{\circ}$ (camera angle 31.75$^{\circ}$), targeting the H$\alpha$ to H$\gamma$ region of the spectrum, and achieving $\sim$850 $< R <$ $\sim$1250. The  integrated exposure time (excluding overheads) was 600~s. The second setup used a grating angle of 12.875$^{\circ}$ (camera angle 25.75$^{\circ}$), targeting the H$\beta$ region and the blue continuum toward roughly 3500~$\AA$ (rest frame), and had an integrated exposure time (excluding overheads) of 540~s. This spectrum had $\sim$700  $<R<$  $\sim$1000.
We used the 2$\times$2 spatial binning, the faint gain, and the slow readout mode for both setups.

\subsection{Continuum variability overview}     

Here, we briefly discuss the continuum light curve overview, presented in Fig.~\ref{fig:obs_fig1}. Section~\ref{section:xmm} covers the fits to the \textit{XMM-Newton} data, which have the highest signal/noise. Based on the results of these fits, we additionally conducted spectral fits for the eROSITA and NICER data to extract the 0.2–5.0 keV fluxes, as illustrated in Fig.~\ref{fig:obs_fig1}. Comprehensive details of these spectral fits can be found in Appendices \ref{appendix:erosita} and \ref{appendix:nicer}.

The X-ray and optical continua display roughly consistent behavior: both show mild increases from early 2020 through early 2021. Both bands experience data gaps, for example, due to Sunblock for ATLAS, from March~2021 through August~2021. In August~2021, both bands are sampled at the highest fluxes we measure (including eRASS4); unfortunately we cannot be certain precisely as to when the true intrinsic peak in any monitored band -- X-ray, UV, optical, or IR -- occurred. Then, both optical and X-ray, along with the UV M2 band, all decrease through early 2022, although we caution that X-ray and UV M2 lack good time sampling here. Both ATLAS bands are relatively monotonic in their decreases.
The X-ray and optical bands are consistent with zero lag (as per visual inspection). Due to the data gaps in both light curves, we cannot establish if any non-zero inter-band lag up to a timescale of very roughly six months exists.

\section{X-ray data analysis}
\label{sec:sec3}

\begin{figure*}
\includegraphics[width=\linewidth]{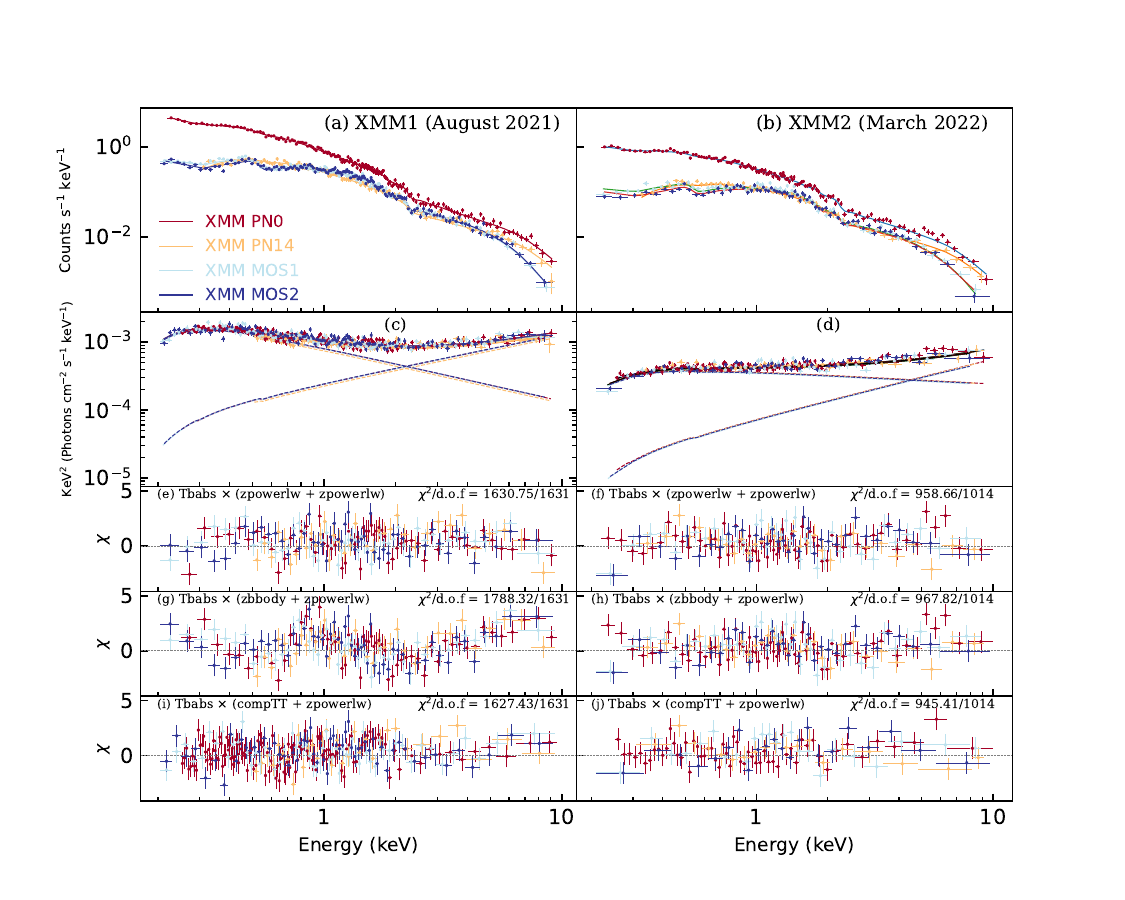}
    \caption{ 
\textit{XMM-Newton} EPIC spectra and best-fitting models.  
The left column shows the data and models for the XMM1 observation in August 2021, taken during the flaring of the event. The right column shows the data and models for the XMM2 observation, taken six months later as the flare subsided. Panels (a) and (b) show the counts spectra, with the best-fitting double power-law model plotted as a solid line. Panels (c) and (d) show the "unfolded" best-fitting  double power-law models and data; the individual power-law components are plotted with dashed lines, and the summed model is shown as a solid line. Panels (e) and (f) show the residuals to double power-law model fits, which are our best-fitting phenomenological models. Panels (g) and (h) show the results from fitting a blackbody component to the soft band plus a hard X-ray power law. Panels (i) and (j) show the results from fitting a Comptonization model (\text{compTT}) to the soft band plus a hard X-ray power law.}
\label{fig:Xray_fig1}    

\end{figure*}

\renewcommand{\arraystretch}{1.3}
\begin{table}
\caption{Results of the spectral fit to the \textit{XMM-Newton} EPIC spectra.}
\centering
\begin{tabular}{llcc}
\hline\hline
Component & Parameters & August 2021 & March 2022\\
&  & XMM1 & XMM2\\
\hline

\multicolumn{4}{c}{ \textsc{constant $\times$ TBabs $\times$ (zpowerlw+zpowerlw) }}                   \\
\hline 
SXPL & $\Gamma_{\rm SX}$ & $2.79 \pm 0.04$    & $2.19^{+0.08}_{-0.06}$   \\
 & Norm.\ ($\times 10^{-4}$)           & $9.59^{+0.52}_{-0.49}$  & $4.17^{+0.35}_{-0.36}$\\
HXPL & $\Gamma_{\rm HX}$  & $1.28^{+0.08}_{-0.10}$     & $1.12^{+0.15}_{-0.18}$ \\
 & Norm.\  ($\times 10^{-4}$)       & $2.67^{+0.43}_{-0.48}$  & $0.79 \pm 0.32$\\
& $\chi^2/d.o.f$ & 1630.75/1631     & 958.66/1014 \\
& $\chi^2_{\rm red}$ & 0.99     & 0.96\\

\hline
\multicolumn{4}{c}{ \textsc{constant $\times$ TBabs $\times$ (zbbody+ zpowerlw)}}                   \\
\hline 
BB & $k_{\rm B}T_{\rm e}$ (keV)              & $0.095 \pm 0.003$      & $0.112^{+0.011}_{-0.015}$ \\
   & Norm.\ ($\times 10^{-5}$) & $2.46 \pm 0.09$      & $0.32^{+0.06}_{-0.08}$ \\
HXPL & $\Gamma_{\rm HX}$                  & $2.03^{+0.03}_{-0.02}$      & $1.84^{+0.03}_{-0.01}$\\
   & Norm.\ ($\times 10^{-3}$) & $1.08 \pm 0.02 $      & $0.46^{+0.02}_{-0.01}$ \\
   & $\chi^2/d.o.f$ & 1788.32/1631     & 967.82/1014 \\  
   & $\chi^2_{\rm red}$ & 1.09    & 0.95\\
\hline
\multicolumn{4}{c}{\textsc{constant $\times$ TBabs $\times$ (compTT+zpowerlw)}}      \\
\hline
compTT& $k_{\rm B}T_{0}$ (eV)      & 20 (F)      & 20 (F)\\
& $k_{\rm B}T_{\rm e}$ (keV)     & 0.3 (F)    & 0.3 (F)\\
& $\tau$       &  $12.3\pm0.3$     & $14.9^{+0.9}_{-1.1}$ \\
& Norm.\ ($\times 10^{-2}$)      & $27.3^{+2.0}_{-1.8}$   & $2.64^{+0.41}_{-0.40}$  \\
HXPL& $\Gamma_{\rm HX}$     & $1.76\pm 0.04$   & $1.67^{+0.06}_{-0.05}$ \\
& Norm.\  ($\times 10^{-4}$) & $7.79^{+0.40}_{-0.39}$    & $3.83^{+0.28}_{-0.27}$ \\
&$\chi^2/d.o.f$      & 1627.43/1631     & 945.41/1014 \\
   & $\chi^2_{\rm red}$ & 1.00    & 0.93\\
\hline

\end{tabular}
\label{tab:obs}

\tablefoot{Hydrogen column density is fixed to the Galactic value of $1.27 \times 10^{20}$~cm$^{-2}$ and the redshift of the source is fixed at $z = 0.057$. (F) means that the parameter is frozen. For power-law components, Norm.\ denotes the value of the power law at 1~keV in units of ph~cm$^{-2}$~s$^{-1}$~keV$^{-1}$.}

\end{table}

The spectral analysis for the X-ray observations are performed using the \texttt{XSpec} (v.12.11.1) package \citep{1996ASPC..101...17A}; we use the $\chi^2$ statistics for the model fitting. The uncertainties on the best-fit parameters are estimated at the 90\% confidence level ($\Delta \chi^{2}= 2.71$ for one free parameter) as derived from Monte Carlo Markov Chains (MCMC) using the Goodman–Weare algorithm. We model Galactic absorption with the total hydrogen column density fixed at $N_{\rm H} = 1.27 \times 10^{20} \rm ~ cm^{-2}$ \citep{2013MNRAS.431..394W}, using the photoelectric cross sections of \citet{1996ApJ...465..487V} and solar abundances from \citet{2000ApJ...542..914W}.

\subsection{XMM-Newton observations}
\label{section:xmm}

We perform combined spectral fitting to the pn0, pn14, MOS1, and MOS2 data for both observations. We use a redshift value $z=0.057$. For both observations, we use the energy range of 0.25--10 keV for pn0 and 0.2--10 keV for MOS1 and MOS2. Since the double-events spectrum has calibration issues below 0.4 keV when using the small window mode, we use the energy range of 0.4--10 keV for pn14. In all models, we apply a constant term to account for cross-instrumental systematics. We left its value for pn0 frozen at unity; values for other instruments were always extremely close to unity.

The X-ray primary continuum of an AGN can be roughly described by a simple power-law model. Hence, we first fit the spectra with a simple Galactic-absorbed redshifted power-law model (\textsc{TBabs} $\times$ \textsc{zpowerlw}) to check for the emission and absorption components in the residual of the fit. This does not give a good fit for either observation, as illustrated by the data-to-model ratios for XMM1 and XMM2 respectively, and suggesting the presence of extra components in the spectra. A soft excess is evident from the positive residuals toward the lower X-ray energies ($\lesssim 1$~keV). One of the best-fitting phenomenological models for both observations was the double power-law model (\textsc{TBabs $\times$ [zpowerlw +zpowerlw]}). In XMM1, the photon index of the power-law fit to the soft X-ray band (SX) was very steep ($\Gamma_{\rm SX} = 2.79 \pm 0.04$), and the photon index of the power-law fit to the hard X-ray component (HX) was quite flat ($\Gamma_{\rm HX} = 1.28^{+0.08}_{-0.10}$). This model had $\chi^2/dof = 1630.75/1631$, with the two model components crossing over at $\sim2$~keV. In XMM2, we find that the soft X-ray component is still present, but has now hardened, as we can fit this component with a power law of photon index $\Gamma_{\rm SX} =2.19^{+0.08}_{-0.06}$; the 0.5--2.0 keV flux of this component has dropped by a factor of 2.5 compared to during XMM1. The hard X-ray component had $\Gamma_{\rm HX} = 1.12^{+0.15}_{-0.18}$. This model had $\chi^2/dof = 958.66/1014$ with the two components crossing over at $\sim 4$ keV. In both spectra, the soft component is very power law-like with no obvious strong curvature and it extends up to roughly 2 and 4 keV in XMM1 and XMM2, respectively. However, it is important to note that it has varied dramatically in both normalization and spectral shape from XMM1 to XMM2. We note that these values of photon index for the soft excess are
similar to those for many other Seyfert 1s' soft
excesses when modeled with simple power laws (e.g., \citealt{2018A&A...611A..59P}).

We check for narrow emission from an Fe-K$\alpha$ line at $E = 6.4$ keV by adding \textsc{zgauss} to the best-fit model, with Gaussian width $\sigma$ fixed to 1~eV. We do not detect the line with high significance in either observation; $\chi^2/dof$ fell by less than 1/1 for both observations. The upper limit on the equivalent width for the Gaussian component using the model (\textsc{TBabs} $\times$ \textsc{[zpowerlw + zpowerlw + zgauss]}) was found to be 45~eV and 155~eV in XMM1 and XMM2, respectively.

Best-fit model parameters for the double power-law model are given in Table~\ref{tab:obs}. We note that the values of the hard X-ray power-law photon index in both observations are unusually hard. It is possible that this very hard continuum is due to a strong contribution from a Compton reflection continuum. However, given the limited bandpass in which to study the hard component, it is not possible to test such a
scenario (although the lack of a very strong Fe K$\alpha$ emission line would argue against it).

\paragraph{Physically motivated models:} We tried both phenomenological and physically motivated models to explain the soft excess. Some of the tested models and the resulting fit values are tabulated in Table~\ref{tab:obs}. We first fit the spectra with a redshifted hard X-ray (HX) power law and a single blackbody for the soft excess. 
For XMM1, this model yielded a poorer fit compared to the double power-law model, with $\chi^2/dof = 1788.32/1631$, and with worse fit
residuals, as illustrated in 
Fig.~\ref{fig:Xray_fig1}(g). For XMM2, fitting a
blackbody also yielded a poorer fit compared to the double power-law
model, although the discrepancy was not as bad as for XMM1. Here, $\chi^2/dof$ was 967.82/1014, and residuals are plotted in Fig.~\ref{fig:Xray_fig1}(h). We also tried fitting the soft excess with \textsc{diskbb} and \textsc{diskpbb} components; both are multiblackbody accretion disk emission with temperature as a function of radius $T(r) \propto r^p$,  where $p$ is fixed to $3/4$ for \textsc{diskbb} and is a free parameter for \textsc{diskpbb}. However, neither component
could fit the soft band in a satisfactory way, yielding poor data/model residuals and
values of $\chi^2/dof > 1.2$ for both. 

Next, we tried the Comptonization model \textsc{compTT}, described in \citet{1994ApJ...434..570T}, for the soft energy band. We fixed the seed photon temperature $T_{0}$ to a value of 20~eV, and the plasma temperature $k_{\rm B}T_{e}$ to a low energy value of 0.3~keV for both observations. This model produced a good fit to both observations, with values of $\chi^2/dof \sim $ 1.00 (XMM1) or 0.93 (XMM2) and data/model residuals virtually identical to those for the double power-law fit. The optical depths were $\tau = 12.3\pm0.3$ (XMM1) and $\tau = 14.9^{+0.9}_{-1,1}$ (XMM2), and with hard X-ray photon indices $\Gamma_{\rm HX} = 1.76 \pm 0.04$ (XMM1) and $\Gamma_{\rm HX} = 1.67^{+0.06}_{-0.05}$ (XMM2).

We also tested if the soft excess could be attributed to relativistically blurred emission from the ionized, inner disk region. We used \textsc{relxill}, with emissivity index frozen to 3, iron abundance frozen to solar, outer radius frozen to 400 $R_{\rm g}$, power-law cutoff frozen to 300 keV, and inclination
angle frozen to $45^{\circ}$. We obtained reasonable fits for both observations: For XMM1 (XMM2), we obtained $\Gamma_{\rm HX} \sim 1.45$ ($\sim$1.17)
and ionization values of log($\xi$[erg cm s$^{-1}$]) = 4.7 (3.7), and values of $\chi^2/dof \sim$ 0.99 (XMM1), 0.94 (XMM2). In both cases, black hole spin
$a^{*}$ pegged at 0.998, as a response to the extreme smoothness of the observed soft excess.
However, in the \texttt{relxill} model, relatively higher disk ionization should correspond to a flatter soft excess (\citealt{2013ApJ...768..146G}), contrary to what is observed.
%%%See e.g, J. Garcia + 2013 ApJ 768 148, Figure 3 %%%%
Consequently, we hold that thermal Comptonization models thus provide the most plausible description of the soft excess.

\begin{figure*}[ht!]
  \centering
%\begin{minipage}[b]{0.49\textwidth}
%\includegraphics[width=0.99\columnwidth]{figures/fig5a.pdf}%[height=7cm, width =9cm]
%\end{minipage}
%\hfill
%\hspace{-3cm}
%\begin{minipage}[b]{0.49\textwidth}
  \includegraphics[width=1.99\columnwidth]{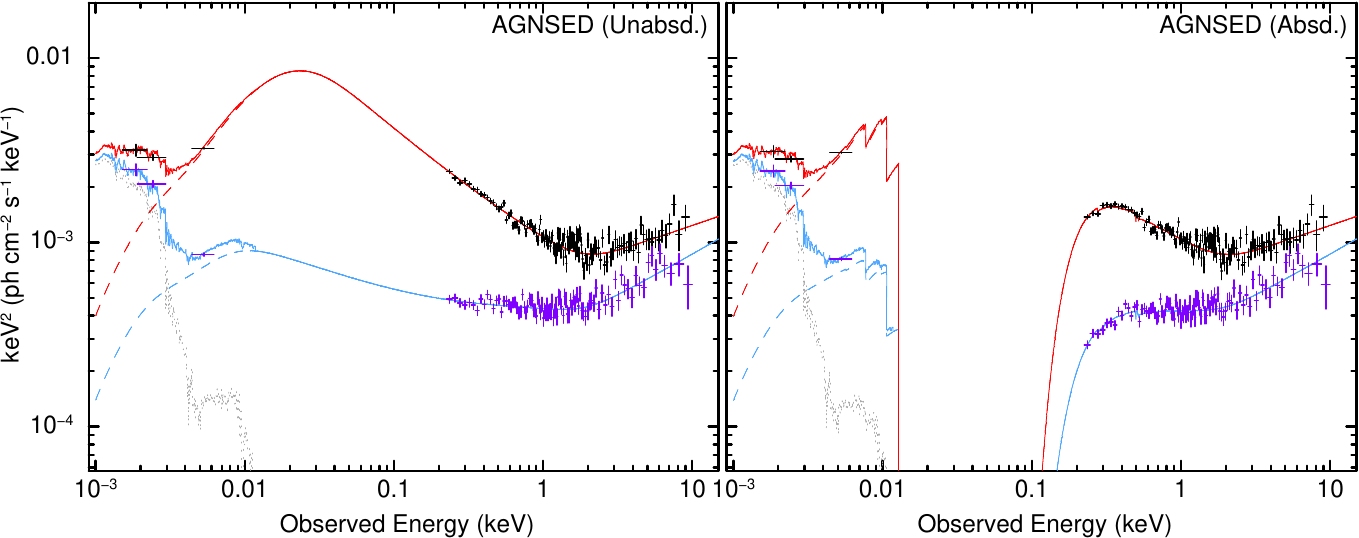}
%\end{minipage}
\caption{     
Flare peak and post-flare SEDs and the best-fitting \textsc{agnsed} model. In the right panel, we present the (observed and absorbed/reddened) data and the best-fitting models. In the left column, we have corrected both data and models for Galactic absorption and reddening. In both panels, black and purple data points denote, respectively, flaring and post-flaring state data. \textit{XMM-Newton} pn data have been rebinned by a factor of 4 for plotting purposes only. In both panels, red and blue lines denote, respectively, the best-fitting models to the flaring and post-flaring states. The dashed lines denote the best-fitting \textsc{agnsed} components, the grey dashed line represents the host galaxy template, and the solid lines represent the total models.   
}
\label{fig:SED1}
\end{figure*}

\section{Modeling the optical to X-ray SED}
\label{sec:sec4}

 \begin{table}
\caption{  \textsc{agnsed} model fits to the flare peak and post-flare SEDs.}
\centering
\resizebox{\columnwidth}{!}{\begin{tabular}{llll}
\hline\hline
Comp. & Par. & August 2021 & March 2022\\
&  & peak & post-flare\\
\hline

\texttt{TBabs} & $N_{\rm H,Gal}$ ($10^{22}$cm$^{-2}$) &  $0.0127$ (F) &  $0.0127$ (F) \\
\texttt{redden} & $E_{B-V_{\rm ,Gal}}$           &  $0.0069$ (F) & $0.0069$ (F)\\
\texttt{redshift} & $z$            & $0.0574$ (F) & $0.0574$ (F) \\
\hline

%\multicolumn{4}{c}{Model 1 -- \texttt{agnsed}}                  % \\
%\hline 
\texttt{agnsed} & $M_{\rm BH}$ ($10^7 \rm M_{\odot}$) &   2.1 (F) & 2.1 (F)\\
 & $\log(\dot{m})$  & $-0.99\pm0.02$ & $-1.38^{+0.11}_{-0.06}$ \\ %%%%%%  & $-0.94 \pm 0.02$    & $-1.39 \pm 0.05$ \\
& $a_{*}$  &   0 (F)  & 0 (F)\\
&  cos$i$  & 0.866 (F)  &    0.866 (F)     \\
& $k_{\rm B}T_{\rm e,hot}$ (keV)   &  $100$ (F)  &  $100$ (F)  \\
& $k_{\rm B}T_{\rm e,warm}$ (keV)   &  $0.33\pm0.04$  &  $0.30\pm0.03$ \\ %%% $0.32^{+0.07}_{-0.05}$  & $0.20 \pm 0.03$ \\
& $\Gamma_{\rm hot}$  & $1.75\pm0.06$ & $1.60^{+0.06}_{-0.08}$ \\ %%% $1.78 \pm 0.08$    & $1.68 \pm 0.05$\\
& $\Gamma_{\rm warm}$  & $2.71\pm0.03$ & $2.35^{+0.04}_{-0.05}$ \\ %%%  $2.74^{+0.03}_{-0.04}$  & $2.39 \pm 0.05$ \\
& $R_{\rm hot}$ ($R_g$) &  $23^{+2}_{-1}$ & $88^{+19}_{-11}$\\ %%%%%  $22^{+2}_{-1}$  & $79^{+9}_{-7}$\\
& $R_{\rm warm}$ ($R_g$) &  $500$ (F)   & $500$ (F) \\
& logrout (\_selfg)  &     $-$1 (F)     &   $-$1 (F)     \\
& H &  10 (F)  & 10 (F)\\
& reprocess &  1 (F)  & 1 (F)\\
& Norm  &   1 (F)  & 1 (F) \\
& $\chi^2/d.o.f$ &  495.10/485 & 339.34/380\\  %% $504.60/491$   & $434.90/385$ \\  
& $\chi^2_{\rm red}$ &  1.02 & 0.90  \\ %%% $1.03$  & $1.13$\\

\hline
%\multicolumn{4}{c}{Model 2 -- \textsc{thcomp*bbody}}                   \\
%\hline 
%\textsc{thcomp1} & $\tau_{\rm warm}$  &   $10.59^{+4.04}_{-12.77}$  & $23.65^{+4.79}_{-1.49}$\\
%& $k_{\rm B}T_{\rm e, warm}$ (keV)   &  $1.58^{+3.21}_{-1.58}$   & $0.50^{+0.99}_{-0.50}$ \\
%& cov. frac.   &  $0.29 \pm 0.02$   & $0.22^{+0.03}_{-0.02}$\\
%\textsc{zbbody} & $k_{\rm B}T_{\rm B}$ (keV)   &  $0.01$ (F)  &  $0.01$ (F) \\
%  & \textcolor{red}{Norm}  &  \textcolor{red}{$(1.57  \pm 0.03) \times 10^{-3}$}   & \textcolor{red}{$(2.86 \pm 0.16) \times 10^{-4}$}\\
%%%%%  & Norm  &  $1.573 \times 10^{-3}$   & $2.86 \times 10^{-4}$\\
%  \textsc{thcomp2} & $\tau_{\rm hot}$  & $4.09^{+2.56}_{-4.09}$    & $2.12^{+0.47}_{-1.46}$  \\
%& $k_{\rm B}T_{\rm e, hot}$ (keV)   & 100.0 (F)     & 100.0 (F) \\
%& cov. frac.   & $6.16^{+4.65}_{-2.54} \times 10^{-3}$     & $4.76^{+2.58}_{-2.27} \times 10^{-2}$  \\
%\textsc{zbbody} & $k_{\rm B}T_{\rm B}$ (keV)   & $0.01$ (F)    &  $0.01$ (F) \\
%  & Norm  &  $(1.57  \pm 0.03) \times 10^{-3}$   & $(2.86 \pm 0.16) \times 10^{-4}$\\
%   & $\chi^2/d.o.f$ &  $632.01/484$    & $425.91/379$ \\  
%   & $\chi^2_{red}$ & 1.31    & 1.12 \\
%\hline
\end{tabular}}

\tablefoot{(F) means that the parameter is frozen.}
\label{tab:SED1}
\end{table}

We construct SEDs for both flare peak and post-flare times, close to eRASS4/XMM1 and eRASS5/XMM2, respectively. However, we caution that it is not clear that eRASS4, XMM1, and the start of the ATLAS window in August 2021 sampled the true luminosity peak of the flare, (what we call 'peak' is the highest luminosity that we measure using \textit{XMM-Newton}; the intrinsic luminosity may have been higher and we missed it). Furthermore, it is not clear that the flare had completely concluded by XMM2 and eRASS5. For simplicity and brevity, we henceforth refer to the two SEDs as "peak" and "post-flare," bearing these caveats in mind. 

We construct both SEDs from \textit{XMM-Newton} EPIC, \textit{XMM-Newton} OM, and ATLAS data. For ATLAS, we use data taken as close as possible to the two \textit{XMM-Newton} observations, which occurred on MJD 59447 (peak) and 59656 (post-flare). For the flare peak, we average data taken within $\pm$8 days of XMM1: c-band:  $810 \pm 18$ $\mu$Jy and o-band: $1150 \pm 24$ $\mu$Jy. For the post-flare SED, ATLAS data only go up to  MJD 59603 (o-band) and 59586 (c-band). We thus took data from the second half of the observing season (after MJD 59500), performed a linear regression, and extrapolated it to MJD 59656. Uncertainties on flux density are estimated from the most recent (closest in time to MJD 59656) cluster of data points: c-band: $581 \pm 17$ $\mu$Jy, o-band: $901 \pm 20$ $\mu$Jy. All SED fits were conducted in \texttt{XSpec}. 

In both SED fits, we account for the host galaxy starlight contribution by using the spectral template of an Sb galaxy from the SWIRE template library \citep{2007ApJ...663...81P} to model it. The Galactic dust reddening is modeled using \textsc{redden} with the value of $E(B-V)$ fixed to $6.86 \times 10^{-3}$ \citep{2011ApJ...737..103S}\footnote{\url{https://irsa.ipac.caltech.edu/applications/DUST/}}. The luminosity distance to the galaxy is 243~Mpc, determined using Ned Wright's cosmology calculator \citep{2006PASP..118.1711W}\footnote{\url{http://www.astro.ucla.edu/~wright/CosmoCalc.html}}; throughout this paper,
we assume a standard flat cosmology with $H_0 = 70$ km s$^{-1}$ Mpc$^{-1}$, $\Omega_{\rm m} = 0.286$, and $\Omega_{\rm vac}=0.714$.

%\subsection{Warm Comptonization with \textsc{agnsed} }
%\label{sec:sed}
For the multiwavelength SED fitting, we applied the \textsc{agnsed} model (\citealt{2018MNRAS.480.1247K}), which includes three emission regions: a standard outer disc region, a warm Comptonizing region, and an inner hot Comptonizing region. In the outer disc region, the emission thermalizes locally, forming a blackbody as in a standard disc, having the modified color temperature-corrected blackbody set to unity, but only in the outer regions, over radii extending from $R_{\rm warm}$ to $R_{\rm out}$. Inward, from $R_{\rm hot}$ to $R_{\rm warm}$, the disk is still optically thick, geometrically thin, but produces a warm Comptonized spectrum to explain the steep UV downturn and the soft X-ray upturn (soft excess) observed in AGN spectra (e.g., \citealt{2007ApJ...668..682D}). The authors consider a truncated disc geometry; interior to $R_{\rm hot}$ and extending down to the innermost stable circular orbit ($R_{\rm ISCO}$) lies the inner hot Comptonization region. The flow does not have an underlying disc component, is geometrically thick, has low optical depth, and produces the primary hard X-ray emission. To model the reprocessing due to the hot corona as an extended source illuminating the warm Comptonization and the outer disc, the model assumes the lamppost geometry to calculate the reprocessed hard X-ray emission by having a spherical hot inner flow from a height $H$ above the black hole, on the spin axis.

We perform a simultaneous fit to both SEDs using the model \textsc{ztbabs $\times$ redden $\times$ (agnsed+swS0template)}. The main parameters of this model are the mass of the SMBH, fixed to $M_{\rm BH} = 2.1 \times 10^7 M_{\odot}$, estimated via the FWHM value of $H\beta$ emission line (discussed in Sect.~\ref{sec:opt}), the distance to the source (fixed to $D = 243$ Mpc), the Eddington ratio (log($\dot{m}$) $\equiv$ log $L_{\rm Bol}/L_{\rm Edd}$), the dimensionless spin parameter of the black hole ($a_*$) fixed to zero, the electron temperature of the hot corona $k_{\rm B}T_{\rm e, hot}$ fixed at the default value of 100 keV, the electron temperature of the warm corona $k_{\rm B}T_{\rm e, warm}$, and the radial size of the hot and warm coronae, $R_{\rm hot}$ and $R_{\rm warm}$ respectively. The log of the outer radius of the disc in units of $R_{\rm g}$ is set to a negative value (logrout$=-1$), to use the self-gravity radius as calculated from \citet{1989MNRAS.238..897L}. We fix the hot corona's scale height to $H=10 R_{g}$ \citep{2017MNRAS.470.3591G}. The inclination angle of the warm Comptonizing region and the outer disc is fixed at $30^{\circ}$. The only parameters allowed to vary between the two SEDs are log($\dot{m}$), $k_{\rm B}T_{e,\rm warm}$, $\Gamma_{\rm hot}$, $R_{\rm hot}$, and $\Gamma_{\rm warm}$; all other parameters are tied between the two SEDs. The model produces best-fitting values of $\chi^2/dof$ = 495.10/485 and 339.34/380 for the flare peak and post-flare SEDs respectively. All best-fit parameter values are given in Table~\ref{tab:SED1}. From the best-fit parameters, we see that the mass accretion rate $\dot{m}$ has decreased from $\sim 0.10$ (flare peak) to $\sim 0.04$ (post-flare), that is, $\dot m $ has decreased by a factor of roughly 2.5. Meanwhile, the radius of the hot corona region ($R_{\rm hot}$) has increased from the flaring state to the post-flaring state by a factor of roughly 4. The best-fitting spectral index slopes and temperature values for the hot and warm Comptonizing region during the flare observation are: $\Gamma_{\rm hot} = 1.75\pm0.06$, $\Gamma_{\rm warm} = 2.71\pm0.03$, and $k_{\rm B}T_{\rm e, warm} =0.33\pm0.04 \rm ~keV$. For the post-flare SED, we obtained $\Gamma_{\rm hot} = 1.60^{+0.06}_{-0.08}$, $\Gamma_{\rm warm} = 2.35^{+0.04}_{-0.05}$, and $k_{\rm B}T_{\rm e, warm} = 0.30\pm0.03$ ~keV. These values of the soft X-ray photon index are consistent with those obtained from the phenomenological double power-law model fits to \textit{XMM-Newton} data in Sect.\ 3.1.
These values of the hard X-ray photon indices are consistent with those obtained from the fits to the \textit{XMM-Newton} data using a model consisting of \textsc{compTT} for the soft band a power law for the hard band. The best-fitting \textsc{agnsed} models are plotted in Fig.~\ref{fig:SED1}.

As discussed below in Sect.~\ref{sec:opt}, we assign an uncertainty of 0.45 dex to the estimate of $M_{\rm BH}$. We thus refit \textsc{agnsed} with values of $M_{\rm BH}$ factors 2.8 higher and lower than above. Values of log($\dot{m}$) are shifted by roughly 0.5--0.6 compared to the $2.1\times10^7 M_{\odot}$ case. Meanwhile, photon indices usually shift by roughly 0.1 to 0.3, and $k_{\rm B}T_{\rm e,warm}$ usually changes by 0.1 or less.  $R_{\rm hot}$ changes by factors ranging from 1.3 to 2.0 depending on mass. However, for all masses tested, the relative changes in best-fitting model parameters between the high- and low-flux states remain robust: log($\dot{m}$) always decreases by 0.35--0.45; $\Gamma_{\rm warm}$ always decreases by roughly 0.4, and $R_{\rm hot}$ always increases by a factor of roughly 3--4 from the high-flux state to the low flux state.

\section{Optical spectroscopic analysis} 
\label{sec:opt}

\begin{figure*}
 
\includegraphics[width=17cm]{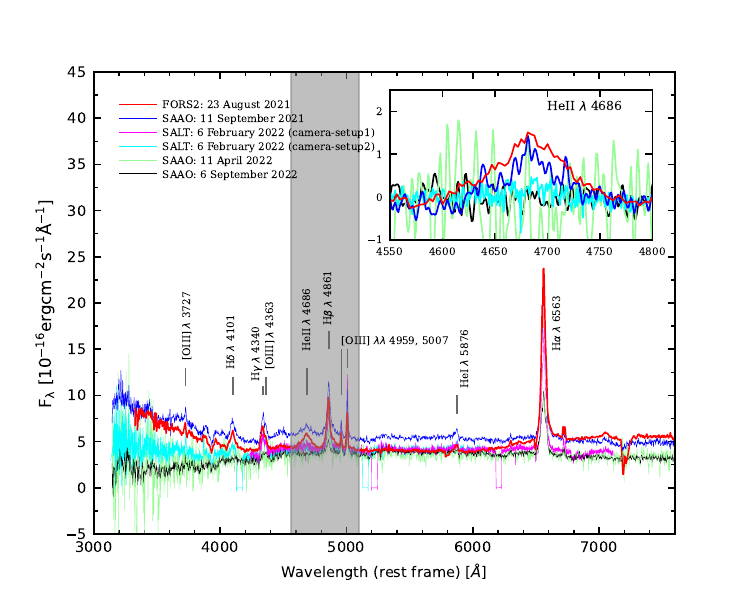}
\caption{Optical spectra of J0408$-$38 taken with VLT, SAAO~1.8m, and SALT. The flux has been re-scaled for all the spectra using the [\ion{O}{iii}] $\lambda$5007 emission line. Prominent emission lines are marked. In the inset figure, we show the window range of 4550--4800 $\AA$; spectra are scaled to have matching continuum levels. The \ion{He}{ii} $\lambda$4686 emission line fades dramatically over the course of the observations. }
\label{fig:Opt_fig1}    
\end{figure*}

 \begin{table*}
\caption{\label{t7} Optical spectral line fit results.}
\centering

\resizebox{\textwidth}{!}{%
\begin{tabular}{llcccccc}
\hline\hline
MJD$^a$ & Instrument &\ion{He}{ii} flux & \ion{He}{ii} FWHM & H$\beta_n$ flux &  H$\beta_n$ FWHM & H$\beta_b$ flux &  H$\beta_b$ FWHM \\
  (d)  & & $10^{-16}~\rm erg$ $\rm cm^{-2}~\rm s^{-1}$& $\rm km$ $\rm s^{-1}$ & $10^{-16}~\rm erg$ $\rm cm^{-2}~\rm s^{-1}$ & $\rm km$ $\rm s^{-1}$ & $10^{-16}~\rm erg$ $\rm cm^{-2}~\rm s^{-1}$ & $\rm km$ $\rm s^{-1}$ \\
\hline

59449 & FORS2 (\#1) & $100.1 \pm 22.6$ & $4573 \pm 153$  & $23.1 \pm 3.5$ &  $889 $ & $87.6 \pm 9.0$ & $2107 \pm 222$ \\
59468 & SAAO ~(\#2) &  $48.1 \pm 34.6$ & $3733 \pm 251$ &  $21.3 \pm 5.3$     & 551   & $155.7 \pm 22.5$ & $2316 \pm 48$\\
59616 &  SALT ~ (\#3) & $< 4.1$ &  $2276 \pm 396$    & $10.7 \pm 4.9$ & 431 & $97.7 \pm 23.4$& $2054 \pm 65$\\
59680 &  SAAO ~(\#4) & -- & -- &  $30.5 \pm 18.2$    & 689  &  --& --\\
59828 &  SAAO ~(\#5) & $< 1.9$& --& $10.4 \pm 7.3$     & $727$ & $36.5 \pm 30.8$ & $2924 \pm 56$\\
\hline

\end{tabular}%

}
\tablefoot{$^a$Date (Modified Julian Day) of observations. The subscripts $n$ and $b$ indicate the narrow and broad components of the emission lines, respectively. The velocity width of the narrow H$\beta$ line was tied to that for the narrow [\ion{O}{iii}] $\lambda\lambda$5007,4959 lines.}
\label{tab:optspc}
\end{table*}

\begin{figure*}[h]
 \centering
\includegraphics[scale=0.45]{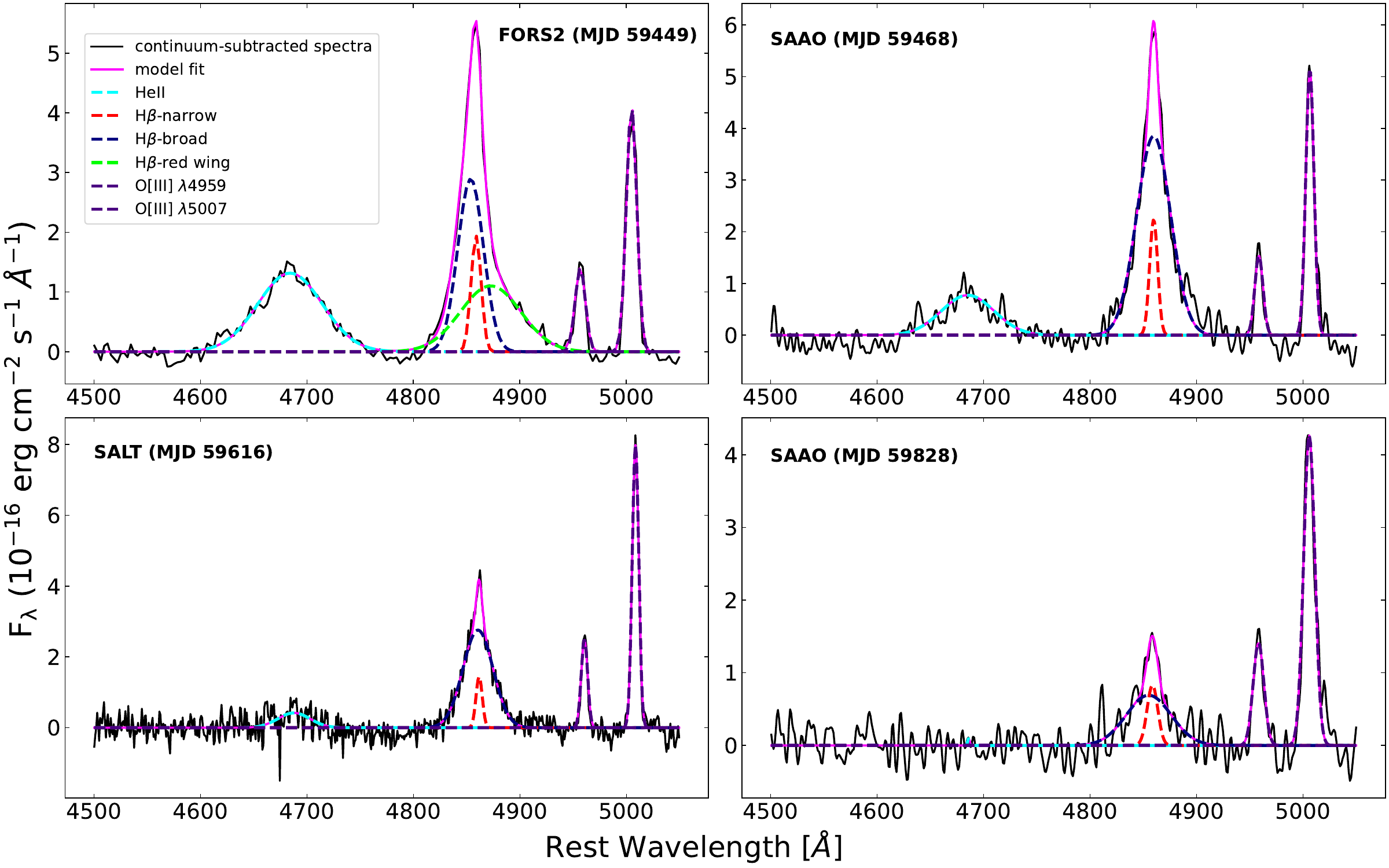}

    \caption[Line fitting of the optical emission line components using \texttt{lmfit}]{The fit residuals after the subtraction of the AGN continuum and host galaxy contribution are represented by the continuous black line.
Here, we model the \ion{He}{ii}~$\lambda$4686, H$\beta$~$\lambda$4861,
and [\ion{O}{iii}]~$\lambda\lambda$5007,4959 emission profiles.  The
magenta line marks the fit of the total spectrum including all
emission components.
}
\label{fig:Opt_fig2}
\end{figure*}

We display our five optical spectra in Fig.~\ref{fig:Opt_fig1}; the most visually prominent line emission features in the 4000--7000~$\AA$ (rest-frame) range include
those associated with H$\alpha$, H$\beta$, H$\gamma$, the [\ion{O}{iii}]~$\lambda\lambda$5007,4959 doublet, and \ion{He}{ii}~$\lambda$4686.
When one is fitting an AGN spectrum, one typically needs to account for the following continuum components: a blue continuum approximated by a simple power law, \ion{Fe}{ii} emission, and the host galaxy. We use the empirical template presented in \citet{2008ApJ...675...83B} for the \ion{Fe}{ii} emission, and an Sb spectrum generated by \citet{2003MNRAS.344.1000B} for the host galaxy template. However, in all our fits, we found no significant contribution from \ion{Fe}{ii} emission and thus excluded this component from our fitting process. 

To account for variations in wavelength resolution and wavelength offsets between different instruments, as well as differing seeing conditions between different observations, 
we scale (grey-shift) each spectrum by fitting the integrated [\ion{O}{iii}]~$\lambda$5007 line flux and wavelength centroid to a reference spectrum following \cite{groningen1992}. 
Specifically, we use the FORS2 spectrum as the reference spectrum to flux scale all the other spectra since it has the highest S/N.
We use the Goodman Weare Markov Chain Monte Carlo algorithm \citep{2010CAMCS...5...65G} for fitting.  We do not correct for different levels of host galaxy contamination that can result from varying slit dimensions. We de-redden the spectra to account for Galactic extinction following \citet{2011ApJ...737..103S}. We performed the spectral analysis using the \texttt{lmfit} python program, which implements the least-squares fitting for the continuum and phenomenological line fitting. The fitting was performed for three spectral windows: 3800--4400 {\AA}, 4400--5050 {\AA}, and 5200--6800 {\AA} in the rest frame. To fit the emission lines in the spectrum, we fit a local continuum and used Gaussian profiles to model the various emission line features. 

We use residuals from the continuum-subtracted spectra to model the emission line features around H$\alpha$ and H$\beta$ in the wavelength ranges of 6450--6700 {\AA} and 4800--4920 {\AA}, respectively. We fit the narrow [\ion{O}{iii}] $\lambda\lambda$5007,4959 lines.  For all other narrow-line components, we fixed their velocity widths to have the same velocity width as [\ion{O}{iii}] $\lambda$5007. 

In the detailed spectral fitting for the window range 4400--5050 {\AA}, we fit \ion{He}{ii} $\lambda 4686$, H$\beta$ $\lambda4861$, and [\ion{O}{iii}] $\lambda\lambda$5007,4959; the    spectra and fits are shown in Fig.~\ref{fig:Opt_fig2}. Table~\ref{tab:optspc} lists our best-fitting model parameters. Of particular note is the flux of the broad \ion{He}{ii} $\lambda$4686 emission line: it fades dramatically over the course of the observations, concurrent to the fading in the X-ray/optical/UV continua (as shown in Fig.~\ref{fig:Opt_fig1}.) We note that the S/N of spectrum \#4, taken at MJD 59679 using the SAAO 1.9m telescope, was too poor to obtain reliable constraints on H$\beta$ profile parameters, and the \ion{He}{ii} $\lambda$4686 line was also not detected within the noise limits. More specifically, the
\ion{He}{ii} line flux decreases from $(100.1 \pm 22.6) \times 10^{-16}$~erg~cm$^{-2}$~s$^{-1}$ to $ < 4.1 \times 10^{-16}$~erg~cm$^{-2}$~s$^{-1}$ from spectrum \#1 to spectrum \#3 and then to being undetected in spectrum \#5, with an upper limit of $1.9 \times 10^{-16}$~erg~cm$^{-2}$~s$^{-1}$.

In addition, our fits to spectrum \#1 
%(VLT, MJD 59449) 
required an extra redshifted broad Gaussian component, as plotted in  Fig.~\ref{fig:Opt_fig2}. The best-fitting energy centroid of this extra broad redshifted component is $4873 \pm 3$ $\AA$, corresponding to a best-fit velocity redshift of $ 710 \pm 200$ km s$^{-1}$.  The width of this component is $\sigma = 28.4 \pm 1.6$ $\AA$, which corresponds to a FWHM velocity of 4118 $\pm$ 232 km s$^{-1}$.  

We consider the possibility that this feature redward of H$\beta$ could be associated with a broad \ion{He}{i} $\lambda$4922 "red shelf", as identified in optical spectra of Seyferts by \citet{2002A&A...384..826V}.
As the ionization potential of \ion{He}{i} is 24.6 eV, it is possible that such a feature would, like
\ion{He}{ii}, respond to a bluer continuum than that powering Balmer lines and thus exhibit stronger variability than that displayed by 
H$\beta$. However, such an identification is tentative at best, due to blending with other lines. Moreover, the best-fit
value of the red wing's centroid, $4873 \pm 3$ $\AA$, would seem to argue against an identification in \ion{He}{i} $\lambda$4922.

Overall, the intensity of the broad H$\beta$ component (excluding the redshifted component detected in spectrum \#1) decreases from spectra \#1 and \#2 to spectrum \#5 (we ignore the low-signal/noise spectrum \#4 in this statement), concurrent with the X-ray, UV, and optical continuum flux decreases. 
More specifically, the H$\beta$ intensity decreases by a factor of $\sim$3 (roughly the same as the UV M2 and X-ray fluxes). However, \ion{He}{ii} displays a much sharper drop: a factor of $\sim$30. In the Discussion (Sect.~\ref{sec:sec6}), we discuss BLR diagnostics based on these measured line and continuum variability behaviors.
 
We can obtain virial estimates for both the radial location of the H$\beta$-emitting region of the BLR, $R_\textrm{BLR}$, and the black hole mass, $M_\textrm{BH}$, using the empirical relation between optical luminosity and BLR radius $R_\textrm{BLR}$ in nearby Seyferts (\citealt{2013ApJ...767..149B}). 
From the optical spectral fits, we find an average flux density of ${\lambda}F_{5100\AA}$ to be $4.18\times10^{-16}$~erg~cm$^{-2}$~s$^{-1}$ $\AA^{-1}$, which for a luminosity distance of 243~Mpc corresponds to a monochromatic luminosity of ${\lambda}L_{5100\AA} = 2.0\times10^{43}$ erg~s$^{-1}$. 
We use the average value of H$\beta$ FWHM velocity across all spectral fits, 2514~km~s$^{-1}$.  We apply the ${\lambda}L_{5100\AA}-R_\textrm{BLR}$ relation of \citet{2013ApJ...767..149B}, log($R_\textrm {BLR}$, lt-dy) = $K$ + ${\alpha}$log( ${\lambda}L_{5100\AA}$/ ($10^{44}$ erg s$^{-1}$)), with their best-fitting values of $K = 1.56$ and $\alpha=0.55$.  
We obtained $R_\textrm{BLR} = 3.7\times10^{16}$~cm = 14.1 lt-days. Assuming a virial factor $f$ of 1.0, we obtained $M_\textrm{BH} = f R_\textrm{BLR} v_\textrm{FWHM}^2 G^{-1} = 2.1\times10^7 M_{\odot}$. We assign an uncertainty of $\sim$0.13 dex to $R_\textrm{BLR}$, $M_\textrm{BH}$, and $L_\textrm{Edd}$, based on the scatter in the ${\lambda}L_{5100\AA}-R_\textrm{BLR}$ relation of \citet{2013ApJ...767..149B}. 
As a caveat, this use
of single-epoch spectroscopy to estimate black
hole mass is dependent on the assumption that the H$\beta$-emitting region is virialized, even during this flaring event.
To be conservative, we assigned an additional systemic uncertainty based on the maximum and minimum optical/UV fluxes being a factor of two more or less than the average flux, which yielded an uncertainty of $\sim$0.17 dex from \citealt{2013ApJ...767..149B} relation. In addition, the scatter in measured values of H$\beta$ FWHM, on the order of $\pm$20\% from the average value, translates into an additional uncertainty of $\sim$0.15 dex. The total uncertainty on $M_\textrm{BH}$ is thus $\pm \sim$0.45 dex.

\section{Discussions}
\label{sec:sec6}

\subsection{Summary of main results}

We summarize our key findings from the X-ray data analysis, the modeling of the optical-to-X-ray SED, and the optical spectroscopic analysis of the flare observed in J0408$-$38 detected by eROSITA below:

\begin{itemize}
    \item The source was observed to be flaring wherein the soft X-ray flux increased within roughly six months by a factor of roughly 5 (compared to eighteen months earlier, the flux increase was a factor of roughly 18). Concurrently, the total optical continuum emission (AGN + host galaxy) increased by$\sim 25$--30\%, as observed via ATLAS photometry.
    
    \item As the source started to fade again, the X-ray decreased by a factor $\sim 3$ over a period of roughly six months. The UV M2 continuum decreased by a factor of 3.8 during this time. As inferred from SED fits, the optical continuum emission concurrently decreased by roughly 2.5.  %% $\sim$4.
    
    \item The spectral shape of the soft X-ray excess was found to be strongly varying between the bright and the faint state of the source. \textit{XMM-Newton} EPIC spectra reveal that the power-law photon index of the soft X-ray band was very steep during the peak of the flare ($\Gamma_{\rm SX} \sim 2.8$) and flattened considerably by the end of the flare six months later ($\Gamma_{\rm SX} \sim 2.2$). Thermal Comptonization models are preferred over ionized disk reflection and blackbody models to describe the soft excess.

    \item We model the broadband optical/UV-X-ray SEDs both near the flare peak, and during post-flare times using the \textsc{agnsed} thermal Comptonization model to describe the broadband emission, including a warm Comptonized component to describe the soft X-ray excess.

    \item The SED modeling implies that from near the flare peak to post-flare times, the bolometric luminosity $L_{\rm Bol}/L_{\rm Edd}$ dropped by a factor of roughly 2.5, from roughly 10\% to 4\%. %% $\sim$3, from $\sim$11 per cent to $\sim$4 per cent. 

    \item From the optical spectral observations, we see that the broad \ion{He}{ii} $\lambda$4686 emission line was present during the flaring state of the source but faded dramatically (a factor of at least 30) over the next few months.

    \item We detect a redshifted (by 11 $\AA$) broad component to H${\beta}$ line emission at flare peak. It disappeared over the successive two months as the continuum faded.  %This red-shifted broad component could be associated with the in-flow of material onto the SMBH.

\end{itemize}

Based on the observations mentioned above, we explored the possible physical scenarios in the sections that follow.
%----------------------------

\subsection{Origin of the soft X-ray excess}

The soft X-ray excess has been identified in some Seyferts to vary independently and sometimes more slowly than the hard X-ray coronal power law (e.g., \citealt{2009ApJ...691..922M}, NGC 3227; \citealt{2001ApJ...561..131T}, Ark 564; \citealt{2002ApJ...568..610E}, Ton S180). The soft excess can even vary independently of the hard X-ray variability on timescales of years (\citealt{2011A&A...534A..39M}, \citealt{2013A&A...549A..73P}). Moreover, strong variations in the strength of the soft X-ray excess have been seen, and it has even been
observed to disappear between observations (\citealt{2009ApJ...705..496M}, \citealt{2012ApJ...759...63R}, \citealt{2018MNRAS.480.3898N}). Below we discuss the flare in the context of several models, with a focus on the strong spectral variation observed in the soft X-ray excess.

In many Seyferts, the soft excess is modeled well by the ``warm'' Comptonization model wherein the optical/UV thermal photons are up-scattered in a Comptonizing medium that is optically thicker ($\tau \sim$ 10--40) and lower in temperature ($k_{\rm B}T_{\rm e} \lesssim 1$ keV) compared to the hot corona, which is responsible for the primary X-ray emission ($\tau \lesssim$ 1--2; $k_{\rm B}T_{\rm e}$ $\sim$ several tens to $\sim100$ keV). 
As one typical example, \citet{2013A&A...549A..73P} used ten simultaneous \textit{XMM-Newton} and INTEGRAL observations of Mrk 509 to find that a hot ($k_{\rm B}T_{\rm e} \sim$ 100 keV), optically thin ($\tau \sim 0.5$) corona is producing the primary hard X-ray continuum and that the soft excess can be modeled well by a warm ($k_{\rm B}T_{\rm e} \sim$ 1~keV), optically thick ($ \tau \sim$ 10--20) plasma.
This model has been successfully applied to many other sources such as 
NGC~5548 \citep{1998MNRAS.301..179M}, 
RE~J1034+396 \citep{2009MNRAS.394..250M}, 
RX~J0136.9$-$3510 \citep{2009MNRAS.398L..16J}, Ark~120 \citep{2014MNRAS.439.3016M}, 
1H~0419$-$577 \citep{2014A&A...563A..95D}, Mkn~530 \citep{2018MNRAS.478.4214E}, 
and Zw229.015 \citep{2019MNRAS.488.4831T}, with warm corona temperatures and optical depths similar to those obtained for J0408$-$38. The Comptonization model is supported by correlations between optical-UV and soft X-ray variability, as observed
for instance by \citet{1996ApJ...470..364E}; \citet{1997ApJ...479..222M}; \citet{2011A&A...534A..36K}. In the case of J0408$-$38, the Compton up-scattering of a burst of optical/UV thermal emission from the disk could have produced a corresponding burst of X-ray emission. This could have effectively cooled the warm corona, leading to a steepening in the soft X-ray spectral slope. As the disk emission fades, the corona %receives an input of heating and/or 
may not cool as effectively (for example, the optical depth could increase), leading to the soft X-ray slope getting flatter again. 

From the best-fitting parameters to our source J0408$-$38 using the \textsc{agnsed} model discussed in Sect.\ 4, we detected a varying soft-excess from the outburst phase to the declining phase with the photon index varying from 
$\Gamma_{\rm warm} = 2.71\pm0.03$ to $2.35^{+0.04}_{-0.05}$. 
The temperature of the warm corona was roughly 0.3~keV in both states. 
It is difficult to state why there would not be major changes in $k_{\rm B}T_{\rm warm}$ given the change in seed photon luminosity on strictly physical grounds given that in general, we do not know the physical origin for the warm corona in Seyferts.
However, in the context of the geometry of the \textsc{agnsed} model, then it is possible that the power dissipated in the warm corona remained roughly constant (\citealt{2018MNRAS.480.1247K}).

We estimated the optical depths of the warm and hot Comptonizing regions using their respective best-fit electron temperature and the X-ray power-law photon-index values using the following equation (\citealt{1996MNRAS.283..193Z}):
 
 \begin{equation} \label{eq:1}
  \tau = \sqrt{\frac{9}{4} + \frac{3}{\theta_{\rm e} \lbrack (\Gamma + \frac{3}{2})^2 - \frac{9}{4} \rbrack}}  - \frac{3}{2},   
 \end{equation}
 
\noindent where $\theta_{\rm e} = k_{\rm B}T_{\rm e}/m_{\rm e}c^2$ and $\Gamma$ is the photon index. The optical depth values are both estimated for the warm corona using Eq.~\ref{eq:1}: (a) peak: $\tau_{\rm hot} \sim 0.5$, $\tau_{\rm warm} \sim 16$; (b) post-flare: $\tau_{\rm hot} \sim 0.6$, $\tau_{\rm warm} \sim 23$. Due to poor spectral constraints, the values from our best fits should not be taken at face value, just qualitatively, but the SEDs seem to be consistent with a moderate increase in $\tau_{\rm warm}$.
All of the inferred parameter changes are associated with, and likely caused by, variations in the Eddington ratio, $L_{\rm Bol}/L_{\rm Edd}$, which decreased from 
10\% during the highest flux state we sampled to 4\% over a period of roughly 6 months.

\subsection{Characteristic variability timescales}
In this section we review the variability timescales associated with a standard Shakura-Sunyaev disk (a reasonable geometry to adopt given the inferred accretion rate); we compare them to the observed timescales to constrain the possible origin of the flaring observed in this source as done in \citet{2018MNRAS.480.3898N}. We calculate the light-crossing time ($t_{\rm lc}$), the fastest possible variability timescale for an isotropically emitting region, the orbital timescale ($t_{\rm orb}$) associated with Kepler's third law of motion, the thermal timescale ($t_{\rm th}$), which governs heating and cooling, and the viscous timescale ($t_{\rm vis}$), which describes the inward radial drift of matter. We used Eq.~1 of \citet{2019ApJ...870..123E} to estimate the radii of peak flux emission for the optical 
(600~nm) and UV (220~nm) bands as
roughly 55 and 160~$R_{\rm g}$, respectively. 
Here, $R_{\rm g}$ is the gravitational radius of the black hole given as $GM_{\rm BH}/c^2$, $G$ is the gravitational constant, and $c$ is the speed of light in vacuum.
For calculating the timescales, we thus adopted a "typical" radial distance of $r \sim 100 R_{\rm g}$. We assume the viscosity parameter to be $\alpha =0.03$ and set $H/R$ (the ratio of the height of the disk to radial distance) to 0.001. The mass of the black hole is $M_{\rm BH}=2 \times 10^7 M_{\odot}$.

\begin{equation}
   t_{\rm lt} \sim r/c \approx 2.6  ~\rm hours.
\end{equation}
\begin{equation}
    t_{\rm orb} \sim 2 \pi \left( \frac{r^3}{GM_{\rm BH}} \right)^{1/2}   \approx 7.4 ~\rm 
 days.
\end{equation}
\begin{equation}
    t_{\rm th} \sim (2 \pi \alpha)^{-1} \times t_{\rm orb} \approx  39 ~\rm  days.
\end{equation}
\begin{equation}
    t_{\rm vis} \sim \left( \frac{H}{R} \right)^{-2} \times t_{\rm th}  \approx   10^5 ~\rm  years.
\end{equation}

 On the simplifying assumption that the highest flux point observed represents the maximum flux state of the outburst, we consider the rise and decay times of the outburst to be $\sim$ 170 and 190 days, respectively, estimated from Fig.\ref{fig:obs_fig1}. The light-crossing timescale is too short and the viscous timescale for a geometrically thin disk is too long to be comparable to the variability timescale observed in this source. The thermal timescale (alternately, the viscous timescale for a thick disk, wherein $H/R$ approaches 1) best describes the flare.
 We note that if the value of $\alpha$ was lower than assumed above by a factor of $\sim$4--5, or if a radius of $\sim$250~$R_{\rm g}$ were adopted, then the thermal timescale would much more closely match the observed rise and fall times.
We discuss below some of the scenarios or disk instability models below that could possibly cause variability at such timescales.

\subsection{Disk instability as the origin of the flare in J0408$-$38}
As noted previously, some accreting black holes display much stronger variability than that associated with the stochastic variability associated with persistent accretion. Instabilities in the accretion disk can potentially produce strong variations in accretion rate, yielding extreme variability in luminosity. We consider several models:

(i) A radiation pressure instability is a candidate for generating flares observed in several Galactic compact sources (e.g., \citealt{1997ApJ...479L.145B}, GRS 1915+105; \citealt{1996ApJ...466L..31C}, GRO J1744$-$28; \citealt{2013MNRAS.436.2334P}, IGR 17091$-$3624), and it is suspected of triggering flares in some Seyfert AGN as well (\citealt{2015ApJ...803L..28G}, IC 3599; \citealt{2015MNRAS.454.2798S}, NGC 3599; \citealt{2019MNRAS.483L..88P}, NGC 1566). The mechanism is outlined in \citet{1974ApJ...187L...1L}. More recently, simulations were conducted by \citet{2020A&A...641A.167S} and \citet{2023A&A...672A..19S}, who model 
an unstable intermediate zone due to radiation a pressure instability operating between the outer cold stable disk and an inner, hot advection-dominated accretion flow. 

As a brief overview, the mechanism involves four stages: a quiescent phase, a rising phase, the outburst phase, and the decay phase. 
Initially, the unstable zone is empty or filled with tenuous gas up to the truncation radius ($\rm R_{\rm trunc}$), beyond which the disk is stable. Gas from the outer disk flows into the inner disk, increasing its surface density, but during this stage, the accretion rate and consequently the luminosity do not increase appreciably. The relevant timescale is the viscous timescale for a geometrically thin disk.
When the radiation pressure exceeds the gas pressure, a heat wave propagates from the innermost disk out to $R_{\rm trunc}$, enhancing the local viscosity, the disk scale height, and local accretion rate. The luminosity increases quickly and so this phase is referred to as the rising phase; the thermal timescale (the time needed to heat the disk) dominates this activity.
An outburst phase follows; emission is dominated by thermal emission from the hot and bright disk. The accretion mass starts to drain, 
and the viscous timescale for a geometrically thick disk dominates the activity.
Then, the slow supply of gas from the outer disk fails to compensate for the pace at which the gas in this region is drained. Hence, the luminous phase fades as the disk is emptied
and the local accretion rate rapidly decreases.
However, the next flare can originate after the inner disk is refilled. Therefore, in this mechanism, recurring flares are possible. The time between successive flares can be approximated using the refilling time of the inner accretion.

In the case of J0408$-$38, the roughly months-long rise and decay times could be attributed to the rising and decay phases of this instability,
in accordance with the thermal timescale (as discussed in Sect.~6.3).
However, our observations may have missed the phase of peak emission, and thus we cannot comment on this phase.
Additional long-term monitoring will be required to see if additional bursts occur, in which case
constraints on the disk refilling time would be attained.

(ii) Another potentially relevant scenario that can cause luminosity flares is the hydrogen-ionization disk instability in the outer accretion disk \citep{2018MNRAS.480.3898N}. The mass accumulated at the circularization radius is initially quite cold, and the temperature is below that which ionizes hydrogen. However, the surface density and temperature build up over time. Hydrogen gets ionized when a temperature of 3000~K is attained; at this stage, photons get trapped by the free electrons, leading to additional ionization of hydrogen. Once all hydrogen gets ionized, the temperature increases rapidly to $10^{4}$~K. Such temperatures enhance the sound speed $c_{\rm s}$ and local accretion rate, causing a heating front to propagate inward.

When the outer disk temperature falls below $10^{4}$~K, a cooling front propagates that turns ionized hydrogen to neutral. However, the reduction of mass accretion is governed by the viscous time of the outer disc and occurs quite slowly. In the case of AGNs, it is plausible for the H-ionization instability to drive the changing-look AGN phenomenon on timescales of years. However, a caveat is that the H-ionization instability tends to operate on timescales of weeks--months in BHXRBs, and the translation of timescales to AGNs in general and to J0408$-$38, in particular, is not clear.

These disk instability models are still under development, and it is not immediately clear which models are best applicable to flaring/changing-state AGN in general, or to J0408$-$38 specifically. Additional observations, including denser monitoring of optical/UV thermal emission to identify the order in which different radial regions of the disk are impacted, may help distinguish among them.

\subsection{On the possibility of a tidal disruption event as the trigger to the flare in J0408$-$38}

Stars passing close to supermassive black holes can get tidally disrupted. For a $ < 10^{8} M_{\odot}$ black hole, the tidal radius lies outside the black hole's event horizon, and the debris produced from stellar disruption can produce emission, primarily in the soft X-ray and ultraviolet bands \citep{1988Natur.333..523R,2011ApJ...735..106D}. Such emissions could decay over timescales of weeks to months. The event rate is estimated to be about $\approx 10^{-5} \text{galaxy}^{-1} \rm ~ yr^{-1}$ \citep{2004ApJ...600..149W}. TDEs in quiescent galaxies tend to have asymmetric light curves with a fast rise and a slow decay, with optical/UV thermal emission from stellar streams usually peaking months before X-ray emission from circularized material appears. This scenario does not seem to apply to J0408$-$38, given the simultaneity of the flaring in all bands. However, there are other cases of TDEs occurring in already-active AGNs. Stellar debris can impact the accretion disk, heating it and thus causing excess thermal emission while also supplying extra material for accretion, temporarily increasing $L_{\rm bol}/L_{\rm Edd}$. Shocks in the disk emanating from debris/disk impacts could also temporarily increase the accretion rate (e.g., \citealt{2019ApJ...881..113C}) and
the optical/UV continuum luminosity. For example, the transient PS16dtm is associated with the nucleus of a Seyfert 1 galaxy 
\citep{2017ApJ...843..106B}. This transient is suggested to originate from a tidal disruption event wherein the accretion debris powers the optical/UV emission and obscures the X-ray-emitting region. \citet{2023A&A...672A.167H} reported on an SMBH transient event detected with eROSITA and Gaia: its optical/X-ray continuum variability properties (fast rise, slow decay) and ultra-soft X-ray spectrum ($\Gamma \sim 5$) categorize it as a TDE. However, the optical spectral properties (broad Balmer lines, strong [\ion{O}{iii}], [\ion{N}{ii}], and [\ion{S}{ii}]) categorize the object as a low-luminosity AGN. A high-redshift example is that of the $z=1.1$ quasar SDSS J014124+010306, wherein variability associated with a TDE is claimed to be separated from the AGN continuum variability (\citealt{2022MNRAS.516L..66Z}). Finally, in the local Seyfert 1ES~1927+654, the X-ray emission disappeared following major changes in the optical spectral type (\citealt{2020ApJ...898L...1R}). After the optical and UV outburst, the power-law component produced in the X-ray spectrum vanished, and the blackbody component appeared. \citet{2020ApJ...898L...1R} hypothesized that the event of tidal disruption of a star by the accreting black hole caused depletion of the inner regions of the accretion flow, affecting the magnetic field that powers the X-ray corona.
Additional examples of AGN in which     TDEs are candidates for fueling
temporary flares are \citet{2015MNRAS.454.2798S},
\citet{2015ApJ...803L..28G}, and \citet{2019MNRAS.483L..88P}.

Consequently, we should examine whether
the multiwavelength behavior observed in J0408$-$38 is consistent with TDE-like accretion in an AGN. Speculatively, the transient broad \ion{He}{ii} emission line and/or the redshifted broad H$\beta$ feature captured during the FORS2 spectrum could have emanated from the tail end of the TDE accretion-like debris falling into the disk. In addition, most
optical/UV-selected TDEs have optical/UV luminosities peaking in the approximate range $\sim 10^{43.5-44.5}$ erg s$^{-1}$ (\citealt{2020SSRv..216..124V}).  In J0408$-$38, the 3--30 eV luminosity peaks around $2\times 10^{44}$ erg s$^{-1}$, which does fall into this range.
  %%%%
However, there are also difficulties in applying this explanation to J0408$-$38: The slow rise in optical/UV continuum emission in the months before the flare could argue against a TDE origin, although we may not have observed the true luminosity peak due to the data gaps, and that slow rise could also be normal stochastic disk variability.
   %%%%
In addition, our measured \ion{He}{II} width during the FORS2 spectrum
is only FWHM $\sim$ 3500 km s$^{-1}$; TDEs' \ion{He}{II} widths tend to
be $\sim 10^4$ km s$^{-1}$ and larger (\citealt{2020SSRv..216..124V}).
    %%% See Figure 8 of Van Velzen+2020
Finally, we did not observe strong Bowen fluorescence lines, for example, \ion{O}{iii} $\lambda$3133 or \ion{N}{iii} $\lambda$4640.
We conclude that TDE-like accretion in J0408$-$38 is unlikely.

\subsection{Alternative models}
There are some alternative models to explain large-amplitude variability at optical to X-ray wavelengths observed in some Seyferts and quasars. Such extreme variability is usually too rapid to be consistent with changes in the global accretion rate associated with the inflow in a standard geometrically thin accretion disc. However, extreme and rapid variations in accretion rate can be associated with a geometrically thick disc supported by magnetic pressure (e.g., \citealt{2019MNRAS.483L..17D}). 
Here, the disk's thickness is decoupled from the thermal properties, and extreme variations in $M_{\odot}$ can occur on $\sim$year timescales.

\subsection{Diagnostics of the broad-line region}

\subsubsection{SED evolution as inferred from \ion{He}{ii} and Balmer emission properties}

Here, we briefly discuss the potential diagnostics of J0408$-$38's BLR from the H$\beta$ $\lambda$4861 and \ion{He}{ii} $\lambda$4686 emission lines.  We focus on potential reasons why the decrease in observed \ion{He}{ii} intensity from spectrum \#1 to spectrum \#5 (factor of at least $\sim$30 as it fades to becoming undetected by spectrum \#5) is so much greater than that for H$\beta$ (factor of $\sim$3) during the same time.

First, we consider the possibility that the divergence in behavior between the two lines is driven by evolution in the extreme UV (EUV) continuum, since hydrogen and \ion{He}{ii} emission are driven by continuum emission above $1$~Ryd (13.6~eV) and $4$~Ryd (54.4~eV), respectively. We hypothesize that as the flare fades, the $>$54.4 eV continuum emission fades more sharply than the continuum emission above 13.6 eV, leading to an optical spectrum that becomes increasingly softer and redder. For example, one possibility is that the $>$13.6 and $>$54.4 eV continua are both dominated by thermal disk emission and that the AGN continuum emission follows a "bluer-when-brighter" and "redder-when-fainter" behavior, as has been suggested to occur in Seyfert AGNs by, for example, \citet{1994A&A...291...74P} and \citet{2004ApJ...601..692V}.

We take the fluxes corresponding to the best-fitting \textsc{agnsed} models from Sect.~4. The 13.6~eV -- 10~keV continuum flux drops from $3.72\times10^{-11}$~erg~cm$^{-2}$~s$^{-1}$ in the high-flux state to $4.54\times10^{-12}$~erg~cm$^{-2}$~s$^{-1}$ in the low-flux state, a drop of 8.2. Meanwhile, the 54.4~eV -- 10~keV continuum drops from $1.76\times10^{-11}$ to $3.97 \times10^{-12}$~erg~cm$^{-2}$~s$^{-1}$, a drop of 4.4. That is, the ratio of $>$54.4~eV to $>$13.6~eV model flux actually increases as the flare subsides. This increase occurs because as one increases in energy from the optical/near-UV range to the EUV range, in the context of our best-fitting \textsc{agnsed} fits, the continuum emission becomes dominated by the Comptonized component above $\sim$ 10~eV as it connects the EUV regime to the soft X-ray excess. It is thus not clear that the evolution in the EUV continuum slope can be responsible for the diverging behavior in the \ion{He}{ii} and H$\beta$ lines.  
However, we must caution that our observations of J0408$-$38 completely lack rest-frame energy coverage between 2185 \AA ~(5.7 eV) and 66 \AA ~(211 eV), and this exercise is consequently a strongly model-dependent extrapolation.

Moreover, such an exercise is valid only in the context of our \textsc{agnsed} model fits.
An alternate explanation is that, given the direct link between \ion{He}{ii} flux and the number of ionizing photons in the far-UV part of the SED (e.g., \citealt{2020MNRAS.494.5917F}), it is likely
that the
$\gtrsim$54.4~eV continuum did in fact vary by a factor of roughly 30 across our campaign, while the $\gtrsim$13.6~eV and X-ray continua each only varied by factors of a few.  In this scenario, the far-UV-emitting region of the disk experienced a spike in luminosity of  at least a factor of 30.
Such a statement may be in conflict with
the more modest change in $L_{\rm bol}/L_{\rm Edd}$ and the far-UV flux indicated by the \textsc{agnsed} fits.
However, our modeling using \textsc{agnsed}
was constrained only by optical, UV (5.4 eV), and soft and hard X-ray constraints; in addition,
we may have missed part of the flare due to insufficient data sampling in the time domain.
However, finding a suitable broadband emission model that describes SED variations
wherein the far-UV varies by a factor of 10 more than the X-ray, mid-UV, and optical bands remains a challenge. One speculative possibility is that a disk instability occurred preferentially concentrated toward the annular region of the disk where the bulk of the far-UV emission is emitted. Following Eqn.~1 of \citet{2019ApJ...870..123E} and Sect.~6.3, this radius is expected to be on the order of a few $R_{\rm g}$, that is, very close to the innermost stable circular orbit.

We also consider that the two emission lines are likely probing different radial regions of the BLR, by a radial factor of at least a few. Many works support that both \ion{He}{ii} emission and H$\beta$ emission arise in regions of mostly virialized, and relatively dust-free gas (e.g., \citealt{2006LNP...693...77P}). However, reverberation-mapping campaigns on several nearby Seyferts have revealed that reverberation lags for \ion{He}{ii} are generally shorter than those for H$\beta$, thus supporting radial ionization stratification in the BLR. For example, \citet{2001A&A...379..125K} and \citet{2012ApJ...744L...4G} found \ion{He}{ii} lags to be shorter than those for H$\beta$ by factors of roughly 6--8 in Mkn~110 and Mkn~335, respectively. 
In addition to empirical support, photoionization calculations support that the emission-weighted radii of these two emission lines can differ by roughly an order of magnitude (\citealt{2004ApJ...606..749K}). In the case of J0408$-$38, 
the FWHM velocities for \ion{He}{ii} and H$\beta$, averaged across all observations, imply that if both lines originate in gas in Keplerian motion, then \ion{He}{ii} originates from a radial location roughly one-half that for H$\beta$. Moreover, the observation of \ion{He}{ii} emission varying much more strongly than H$\beta$ emission could, speculatively, indicate diverging physical conditions (geometry, density, or illumination) between the two radially distinct emission regions. As a purely speculative example, there could conceivably have existed some temporary, vertically extended structure in the inner, highly ionized region of the BLR only. Such a structure could have created a larger covering factor for the \ion{He}{ii}-emitting radii than for the larger radii where H$\beta$ is created, as it is most prominent during optical spectrum \#1.

%   change in He2 FWHM width from spec 1 to 2 -- might this be a clue???

Finally, however, a simpler explanation that does not require different covering factors at different radii is based on \ion{He}{ii} and H$\beta$ emission having divergent "responsivities." Following \citet{2004ApJ...606..749K}, responsivity is defined qualitatively as how much a given line's intensity varies for a given variation in ionizing continuum flux. It depends on the spectral shape of the ionization continuum, and on the line-emitting gas' ionization parameter, density, and radial distance from the central engine (additional details in \citealt{2020MNRAS.494.5917F}).  It is defined quantitatively as $\eta \equiv {\Delta} \textrm{log}(F_{\rm line}) / {\Delta}\textrm{log}(\Phi)$, where $F_{\rm line}$ denotes line intensity and $\Phi$ denotes the number of ionizing photons \citep[Eqns.\ 1 and 2 in][]{2004ApJ...606..749K}).

Responsivities generally increase toward larger radii for a given emission line. However, different atomic species can yield vastly different responsivities, 
depending on local conditions such as density, and on how variable the input continuum is. Consequently, there can exist physical conditions where \ion{He}{ii}'s responsivity is substantially stronger than that for H$\beta$.
Multiwavelength observations comparing the variability of \ion{He}{ii} and optical fluxes
are somewhat rare. In one recent example, however, multidecade monitoring of Mkn~110 by \citet{2023MNRAS.519.1745H}
revealed the \ion{He}{ii} flux to vary by $\sim$40 while the optical continuum varied by
a factor of only a few, similar to variability factors observed during J0408$-$38's flare. 
In addition, \citet{2023MNRAS.519.1745H} found line responsivities to evolve between observing epochs, illustrating the extremely dynamic nature of the BLR and the difficulties in finding suitable models.

In our campaign on J0408$-$38, we found that from spectrum \#1 to spectrum \#5, the hydrogen-ionizing continuum drops by a factor of roughly 8 (0.9 in the log), considering the 13.6~eV -- 10~keV continuum emission fluxes from the best-fitting \textsc{agnsed} model fits to the broadband SED. Meanwhile, H$\beta$ intensity drops by roughly 3 (0.5 in the log), yielding $\eta \sim 0.5$. Comparison to Fig.~4 of \citet{2004ApJ...606..749K} indicates that such responsivity is typical for BLR number densities (log($n$, cm$^{-3}$) $\gtrsim$ 10). \ion{He}{ii} intensity drops by at least 30 (1.5 in the log), yielding $\eta \geq$ $\sim$1.7. Such responsivity in \ion{He}{ii} is reasonably achieved toward relatively higher values of $n$, for example, log($n$, cm$^{-3}$) $\sim$ 11--12.

\subsubsection{On the transient red wing to broad H$\beta$ emission}
Finally, we briefly discuss the broad, redshifted emission component detected in the H$\beta$ profile of spectrum \#1 only. This feature can be modeled with a Gaussian with an energy centroid suggesting gas moving away from us with a bulk velocity component along the line of sight on the order of $710$~km~s$^{-1}$. Such a red wing qualitatively resembles those observed in the H$\beta$ profiles of some nearby reverberation-mapped Seyferts, including Arp~151 (\citealt{2014MNRAS.445.3073P}) and 3C~120 (\citealt{2013ApJ...764...47G}). It is not clear if there is any direct connection between J0408$-$38's continuum flare and the (apparently transient) presence of this kinematic component, or if the two are merely coincidental.  
It is also not clear if this red wing arises from a kinematic component that is physically separate and distinct from the gas that yields the (nonshifted) broad emission lines. 
However, the transient nature of this red wing combined with the relative stability of the other broad H$\beta$ component ("broad component 1") seems to favor a component associated with bulk radial motion with respect to the rest frame of the BLR.  
Finally, it is not clear if this feature indicates that the BLR itself contains a kinematic component associated with bulk infall, as measured in several nearby reverberation-mapped Seyferts (\citealt{2013ApJ...764...47G}; \citealt{2014MNRAS.445.3073P}). For face-on
disks, a redshift can indicate infall.
However, for highly inclined disks and for certain viewing angles, a localized outflow driven radially outward could be redshifted.

The broad redshifted kinematic component could, speculatively, be associated with a "Failed Radiatively Accelerated Dusty Outflow" (FRADO; \citealt{2011A&A...525L...8C}; \citealt{2021ApJ...920...30N}). In this model, the BLR is a dusty wind flowing upward from the disk and pushed radially outward by radiation pressure. However, after the gas is heated by the radiation to the point where dust sublimates, the driving force subsides, and the cloud falls ballistically back down to the accretion disk, forming a "failed" wind. The dynamical simulations of \citet{2021ApJ...920...30N} predict that wind heights and velocities depend on source accretion rate, but infall velocities of hundreds to thousands of km~s$^{-1}$ are plausible. In J0408$-$38, the energy centroid of the broad redshifted component suggests an velocity component along out line of sight of $\sim 710$~km~s$^{-1}$. It is (speculatively) possible that the FORS2 spectrum (\#1) caught emission from such a ballistic component, which then faded or impacted into the accretion disk before spectrum \#2 was taken, 19 days later.

\section{Conclusions}
\label{sec:sec7}

The window into time domain X-ray astronomy of eROSITA enables identification of new transient extragalactic X-ray events associated with major changes in accretion rates in AGNs. When multiwavelength follow-up programs are employed, one can track the evolution of individual X-ray spectral emission components, the optical/UV-emitting accretion disk, and the BLR in response to significant changes in accretion rate and luminosity. 

In this work, we have presented a study on a Seyfert 1 AGN, J0408$-$38 that underwent a flaring event detected using eROSITA. We triggered multiwavelength follow-up observations with \textit{XMM-Newton} (EPIC and OM) and NICER near the flare peak and during the flare's fading. We also obtained optical spectroscopic monitoring using VLT-FORS2, SALT-RSS, and SAAO~1.8m-SpUpNIC. The source also flared in the optical and infrared bands, as determined by public photometry from ATLAS and WISE’s NEOWISE-R scans and from photometry taken at CTIO. 
The X-ray flux increased by a factor of approximately five over six months (and a factor of 18 over 18 months) and  had decreased by a factor of about three over the following six months. Meanwhile, the UV flux faded by a factor of almost four from flare peak to post-flare times.

We modeled and compared the X-ray spectra taken during the outburst and during the declining flux state of the source, and we found significant spectral variability in the soft X-ray excess. 
We argue against simple blackbody and ionized disk reflection models in describing the soft excess. A simple power-law model was found to be the best phenomenological fit for the soft excess. Its photon index flattened from $\Gamma_{\rm SX} = 2.79 \pm 0.04$ at the flare peak to $\Gamma_{\rm SX} = 2.19^{+0.08}_{-0.06}$ toward the end of the flare. However, modeling the soft excess via a more physical, warm Comptonizing corona also yielded a good fit, with optical depth increasing slightly (for a fixed electron temperature) as the flare faded.

We demonstrated that the broadband SED of J0408$-$38, from the optical band to the X-rays (and including the soft X-ray excess), could be modeled with the thermal Comptonization model \textsc{agnsed} such that the soft excess arises in warm Comptonized emission. From the best-fitting parameters, the radial distance of the hot corona region increased from 
$R_{\rm hot} \sim 23$ to $\sim 88 R_{\rm g}$ from the outburst to the declining flux state. Meanwhile, the optical depth increased slightly from $\sim16$ to $\sim23$. In the context of this model, the multiwavelength flaring was the result of a sudden change in the accretion rate $\dot{m}$, that is, the accretion rate decreased by a factor of $\sim$2.5 between the two \textit{XMM-Newton} observations, as the UV and optical fluxes tracked the disk output. The increase in disk luminosity drove an increase in the luminosities of both the hot and warm X-ray coronae. As the flare faded, the optical depth increased, leading to strong spectral variability in the soft X-ray power-law slope.

Our optical spectroscopic monitoring revealed a strongly dimming broad \ion{He}{ii} $\lambda$4686 line. As the optical/UV/X-ray continuum fluxes and the flux of the broad H$\beta$ emission line each decreased by factors of a few, the \ion{He}{ii} flux decreased by a factor of at least 30. Generally, \ion{He}{ii} is expected to originate closer to the central engine than H$\beta$, and the diverging flux behavior observed in J0408$-$38 could, speculatively, indicate diverging physical conditions between the two line-emitting regions -- for instance, geometry (covering factor) or illumination. An alternate explanation, however, could be that the \ion{He}{ii} and H$\beta$ emission have divergent responsivities. A high value of the BLR cloud density could yield a comparatively stronger response in \ion{He}{ii} than for H$\beta$ for a given variation in the ionization continuum flux.

We also observed a broad redshifted emission component to H$\beta$ in the optical spectrum taken during the flaring state (spectrum \#1; VLT-FORS2). Its energy centroid suggests a line of sight bulk velocity on the order of 710~km~s$^{-1}$ moving away from us and may represent a transient
kinematic component that is physically distinct from the BLR clouds that produce the broad, nonshifted Balmer line emission.
It is not clear if the bulk flow indicates an inflow or outflow relative to the BLR given the inherent uncertainties in the geometry, viewing angle, etc. However, such a transient component  could speculatively be associated with a "failed radiatively accelerated dusty outflow"; \citep[FRADO;][]{2021ApJ...920...30N}. In this scenario, the VLT-FORS2 spectrum caught emission from such a ballistic component that then faded or impacted into the accretion disk before spectrum \#2 was taken, 19 days later.

As to the underlying mechanism of the flare, the timescales for the increase and decrease of the flare are broadly consistent with those that could be produced by an accretion disk instability, such as the radiation-pressure instability. 
The timescales for the rising and falling times of the flare are broadly consistent with the thermal timescale near the optical continuum-emitting region of the disk or the viscous timescale for a geometrically thick disk (which approaches the thermal timescale). However, the extreme variation in \ion{He}{ii} flux could indicate that the far-UV continuum did in fact experience a luminosity spike a factor of at least approximately ten stronger than the optical/UV and X-ray regions of the SED. One speculative possibility is that a disk instability preferentially affected the far-UV emitting region of the accretion disk (near the innermost circular stable orbit).

Future campaigns -- either on J0408$-$38 in the case that another flare occurs, or on other flaring Seyferts -- require optical/UV/X-ray photometric monitoring with a sufficiently dense cadence to potentially identify any inter-band lags or leads. Such a result could place constraints on the specific flaring mechanism, disk properties, and the radial direction of propagating fronts.

\begin{acknowledgements}

The authors would like to thank Matteo Guainazzi for the thorough reading and for providing valuable comments that helped enhance the manuscript. The authors also thank Chris Done
for useful scientific discussions.
SK, AM, and TS acknowledge partial support from Narodowe Centrum Nauki (NCN) grants 2016/23/B/ST9/03123 and 2018/31/G/ST9/03224. AM also acknowledges partial support from NCN grant 2019/35/B/ST9/03944. DH acknowledges support from DLR grant FKZ 50 OR 2003. MK is supported by DLR grant FKZ 50 OR 2307. SH is supported by the German Science Foundation (DFG grant number WI 1860/14-1). This work is based on data from eROSITA, the soft X-ray instrument aboard SRG, a joint Russian-German science mission supported by the Russian Space Agency (Roskosmos), in the interests of the Russian Academy of Sciences represented by its Space Research Institute (IKI), and the Deutsches Zentrum f\"{u}r Luft und Raumfahrt (DLR). The SRG spacecraft was built by Lavochkin Association (NPOL) and its subcontractors and is operated by NPOL with support from the Max Planck Institute for Extraterrestrial Physics (MPE). The development and construction of the eROSITA X-ray instrument were led by MPE, with contributions from the Dr. Karl Remeis Observatory Bamberg \& ECAP (FAU Erlangen-Nuernberg), the University of Hamburg Observatory, the Leibniz Institute for Astrophysics Potsdam (AIP), and the Institute for Astronomy and Astrophysics of the University of T\"{u}bingen, with the support of DLR and the Max Planck Society. The Argelander Institute for Astronomy of the University of Bonn and the Ludwig Maximilians Universit\"{a}t Munich also participated in the science preparation for eROSITA. The eROSITA data shown here were processed using the eSASS/NRTA software system developed by the German eROSITA consortium. Some of the observations reported in this work were obtained with the Southern African Large Telescope (SALT) under programs 2020-2-MLT-008 and 2021-2-MLT-003. Polish participation in SALT is funded by grant No.\ MEiN nr 2021/WK/01. This research has made use of data and/or software provided by the High Energy Astrophysics Science Archive Research Center (HEASARC), which is a service of the Astrophysics Science Division at NASA/GSFC. 
The Skynet Robotic Telescope Network is supported by the National
Science Foundation, the Department of Defense, the North Carolina
Space Grant Consortium, and the Mount Cuba Astronomical Foundation.
This publication makes use of data products from the Wide-field
Infrared Survey Explorer, which is a joint project of the University
of California, Los Angeles, and the Jet Propulsion
Laboratory/California Institute of Technology, funded by the National
Aeronautics and Space Administration.  This publication also makes use
of data products from NEOWISE, which is a project of the Jet
Propulsion Laboratory/California Institute of Technology, funded by
the Planetary Science Division of the National Aeronautics and Space
Administration.

\end{acknowledgements}

% WARNING
%-------------------------------------------------------------------
% Please note that we have included the references to the file aa.dem in
% order to compile it, but we ask you to:
%
% - use BibTeX with the regular commands:
%   \bibliographystyle{aa} % style aa.bst
%   \bibliography{Yourfile} % your references Yourfile.bib
%
% - join the .bib files when you upload your source files
%-------------------------------------------------------------------

   \bibliographystyle{aa} % style aa.bst
   \bibliography{references.bib}

\begin{appendix}
    
\section{eROSITA observations} 
%\section*{Appendix A: eROSITA observations} 
\label{appendix:erosita}

\renewcommand*\thetable{A.\arabic{table}}
\begin{table*}
\centering
\caption[]{Model fits to the five eRASS spectra.}
\begin{tabular}{llccccc}
\hline\hline
Component & Parameters & eRASS1 & eRASS2 & eRASS3 &  eRASS4 &  eRASS5 \\  \hline
\multicolumn{6}{c}{\textsc{TBabs $\times$ zpowerlw}}                   \\  
\hline

single PL & $\Gamma$ & $2.22^{+0.53}_{-0.51}$ & $2.15\pm0.23$  & $1.87\pm0.27$ & $2.27\pm0.11$ & $1.82\pm0.17$\\
 & Norm.\ $ \times 10^{-4}$  &  $0.73\pm0.17$ & $3.00\pm0.36$  &  $2.73\pm0.37$  & $12.86\pm0.82$  &  $4.45\pm0.48$ \\
 &  $\chi^2/d.o.f$ & 3.24/4 &  19.26/15    & 14.36/10 & 65.81/54 &  17.24/16\\
 &  $\chi^2_{\rm red}$ & 0.81 & 1.28 &  1.44    & 1.22  &  1.08\\
\hline
\end{tabular}
\tablefoot{Norm.\ refers to the 1 keV normalization of the illuminating X-ray
power law, in units of ph cm$^{-2}$ s$^{-1}$ keV$^{-1}$.}
\label{tab:erasstab}

%\tablefoot{Hydrogen column density is fixed to the Galactic value of $1.27 \times 10^{20}$ cm$^{-2}$ and the redshift of the source is fixed at $z = 0.0574$.}

\end{table*}

We bin each eRASS spectrum to a minimum of 15 cts bin$^{-1}$ and modeled them using $\chi^2$ statistics. All spectra are fit moderately well by simple steep power laws due to the limited photon counts (due to the low exposure times), over the energy range of 0.2--3.0 keV. Results are presented in Table~\ref{tab:erasstab}. These spectra are consistent with being completely dominated by the soft excess observed in the \textit{XMM-Newton} EPIC spectra. In Fig.~\ref{fig:Xray_fig2}, we present the data,  best-fitting model, and fit residuals. 
Values of photon indices are consistent with those obtained using \textit{XMM-Newton} and NICER.

\renewcommand*\thefigure{A.\arabic{figure}}
\begin{figure}[H]
\includegraphics[width=\linewidth]{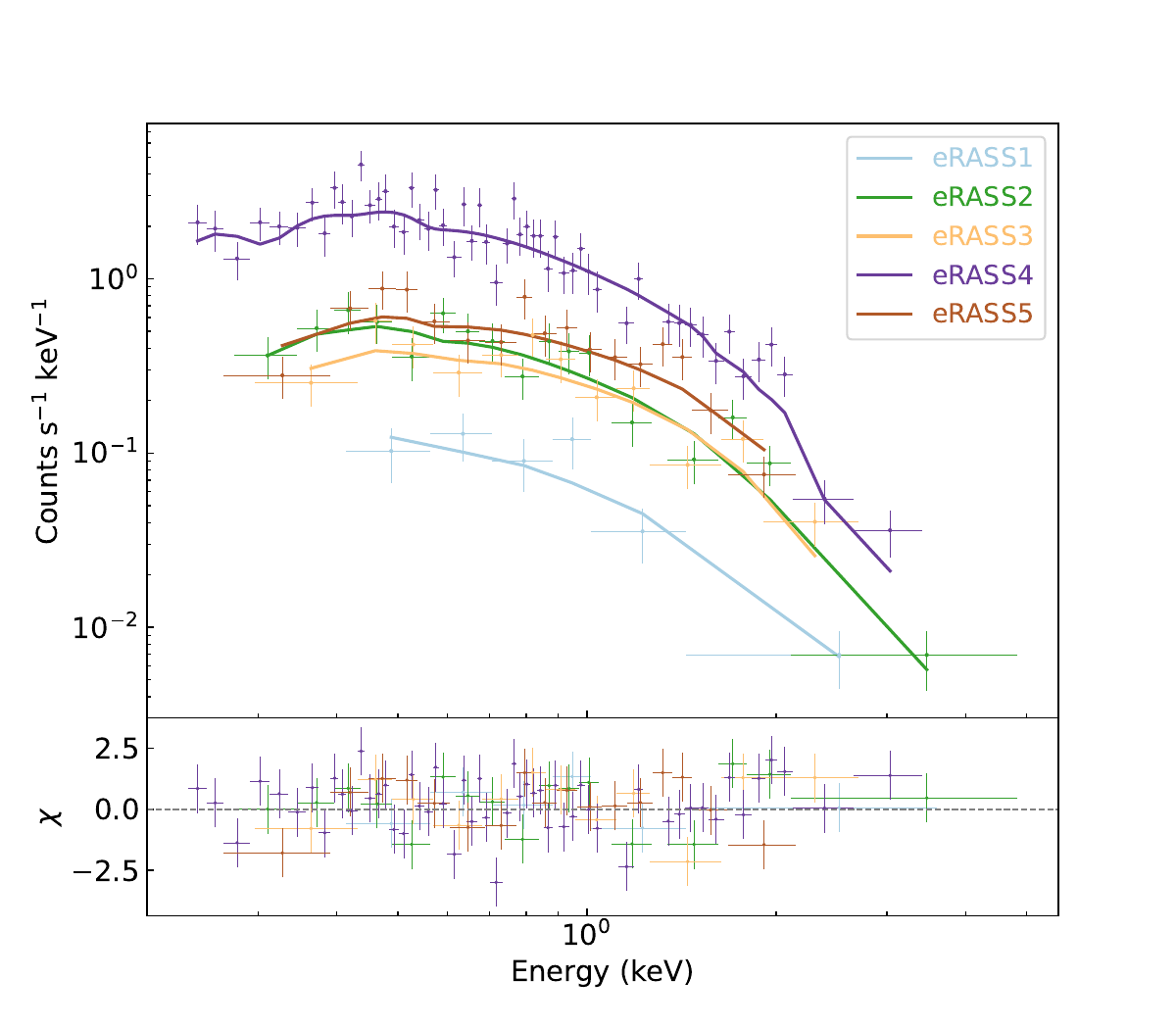}
    \caption{Five eRASS spectra, fit with a simple power-law model (top panel), along with the residuals (bottom panel).}
\label{fig:Xray_fig2}     

\end{figure}

\section{NICER observations } 
%\subsection{NICER observations}
\label{appendix:nicer}

We fit the August 2021 (NICER1) and November 2021 (NICER2) spectra separately; the data are shown in Fig.~\ref{fig:nicer}.  We first applied a simple power law, but in each case, we obtained values of $\chi^2/dof$ above 1.2, with broadband concave data/model residuals similar to those seen in the EPIC data when a single power law was applied. For each spectrum, we instead obtained an excellent fit with a double power-law model, qualitatively consistent with the EPIC fits.  Results are summarized in Table~\ref{tab:obsnicer}.  We do not include any systematic uncertainties associated with incomplete background modeling; typically, for background systematics of $\pm$2\%, $\Gamma_{\rm SX}$, redshifted power-law flux $f_{0.5-2}$, and $f_{2-5}$ are shifted by up to 0.01, $0.03\times10^{-12}$~erg~cm$^{-2}$~s$^{-1}$, and $<$0.01$\times10^{-12}$~erg~cm$^{-2}$~s$^{-1}$, respectively. $\Gamma_{\rm HX}$ is shifted by up to 0.03 (August 2021) or 0.10 (November 2021).

We note that the 0.2--5.0 keV flux measured from NICER1 using the double power-law model fit ($\chi^2/dof = 275.54/281$) extrapolated from 0.4 down to 0.2 keV is $f_{0.2-5.0}= 5.83^{+0.08}_{-0.06} \times 10^{-12} \rm ~erg \rm ~cm^{-2} \rm ~ s^{-1}$. The soft and hard power-law photon indices are $\Gamma_{\rm SX}=2.78^{+0.48}_{-0.27} $  and $\Gamma_{\rm HX}=1.49^{+0.31}_{-0.49} $ respectively. These values are roughly consistent with the fit values from XMM1.  In addition, the flux from NICER2 using the double power-law model fit ($\chi^2/dof = 485.88/430$) is $f_{0.2-5.0}= 3.99^{+0.02}_{-0.25} \times 10^{-12} \rm ~erg \rm ~cm^{-2} \rm ~ s^{-1}$ and is roughly consistent with the interpolation of fluxes between XMM1 and XMM2. The soft and hard power-law photon indices during NICER2 are $\Gamma_{\rm SX}=2.27^{+0.42}_{-0.12} $ and $\Gamma_{\rm HX}=1.08^{+0.63}_{-0.92} $, respectively.

\renewcommand*\thetable{B.\arabic{table}}
\begin{table}[H]
\caption{Model fits to NICER spectra.}
\centering
\begin{tabular}{llcc}
\hline\hline
Component & Parameters & Aug.\ 2021 & Nov.\ 2021 \\
&  & NICER1 & NICER2\\
\hline

\multicolumn{4}{c}{ \textsc{TBabs $\times$ zpowerlw}}                   \\
\hline 
single PL & $\Gamma$ & $2.22 \pm 0.02$    & $2.07 \pm 0.02$   \\
 & Norm.\  ($\times 10^{-3}$)           & $1.31 \pm 0.02$  & $0.93 \pm 0.01$\\
& $\chi^2/d.o.f$ & 363.97/283     & 531.44/432 \\
& $\chi^2_{\rm red}$ & 1.29     & 1.23\\

\hline
\multicolumn{4}{c}{ \textsc{TBabs $\times$ (zpowerlw + zpowerlw  )}}                   \\
\hline 
SXPL & $\Gamma_{\rm SX}$ & $2.78^{+0.48}_{-0.27} $    & $2.27^{+0.42}_{-0.12} $   \\
 & Norm.\  ($\times 10^{-4}$)           & $7.29^{+3.19}_{-3.75}$  & $7.68^{+1.20}_{-4.33}$\\
HXPL & $\Gamma_{\rm HX}$  & $1.49^{+0.31}_{-0.49} $    & $1.08^{+0.63}_{-0.92} $ \\
 & Norm.\  ($\times 10^{-4}$)         & $5.07^{+3.74}_{-3.09}$  & $1.36^{+4.29}_{-1.13}$\\
& $\chi^2/d.o.f$ & 275.54/281     & 485.88/430 \\
& $\chi^2_{\rm red}$ & 0.99     & 0.96\\

\hline

\end{tabular}
\label{tab:obsnicer}

%\tablefoot{Hydrogen column density is fixed to the Galactic value of $1.27 \times 10^{20}$ cm$^{-2}$ and the redshift of the source is fixed at $z = 0.0574$.}

\end{table}

\renewcommand*\thefigure{B.\arabic{figure}}
 \begin{figure*}[h]
\centering
\includegraphics[width=\linewidth]{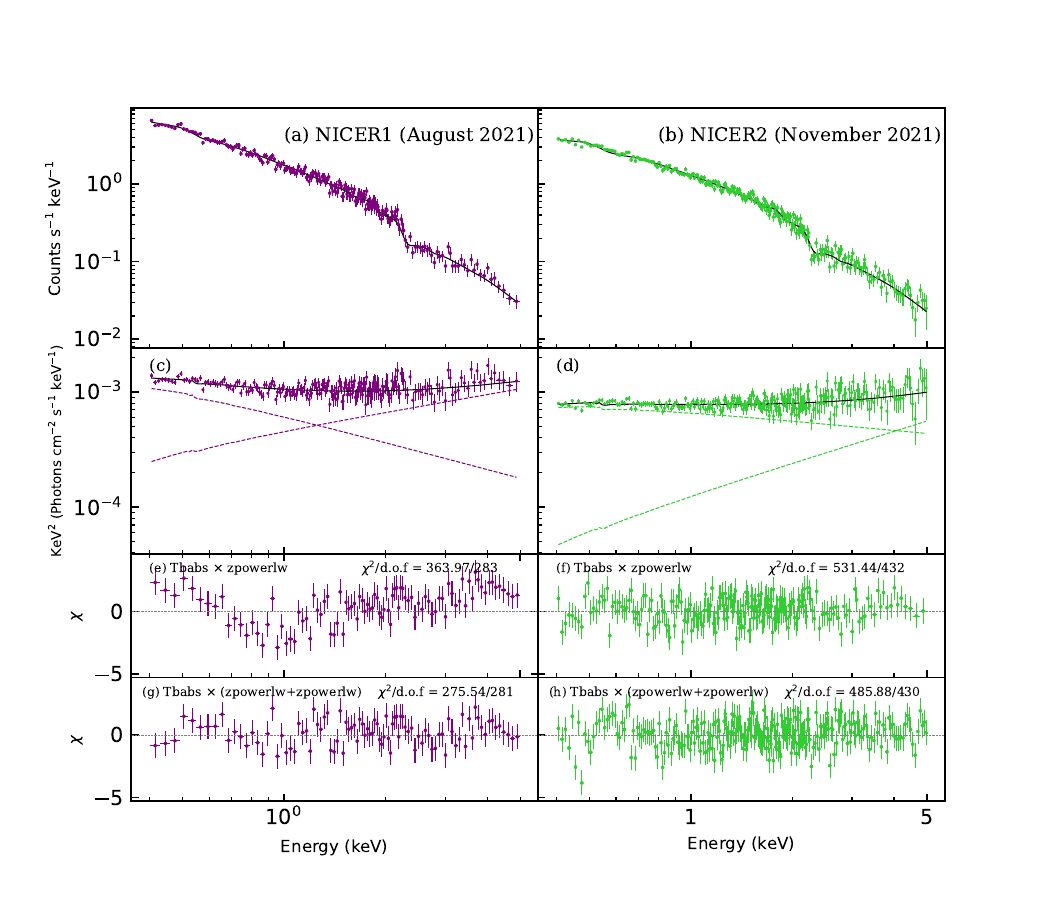}
        \caption[Follow-up NICER X-ray spectrum of J0408$-$38]{NICER X-ray spectra of J0408$-$38  and model fits. The left column shows the data and models for the NICER1 observation in August 2021, taken during the flaring of the event. The right column shows the data and models for the NICER2 observation in November 2021, during the fading of the flare.  Panels (a) and (b) show the counts spectra, with the best-fitting double power-law model plotted as a solid line. Panels (c) and (d) show the "unfolded" best-fitting  double power-law models and data; the 
        individual power-law components are plotted with dashed lines, and the summed model is shown as a solid line.
        Panels (e) and (f) show residuals to a simple power-law fit; the fits are poor and suggest the presence of additional components in the spectra. Panels (g) and (h) show residuals from the double power-law model; the fits are good, and in qualitative agreement with fits to the \textit{XMM-Newton} EPIC spectra.}
\label{fig:nicer}    
\end{figure*}

\end{appendix}

\end{document}